\documentclass[fleqn,usenatbib]{mnras}

\usepackage{newtxtext,newtxmath,cuted,lipsum}
\usepackage[T1]{fontenc} 
\usepackage{graphicx}	
\usepackage{amsmath}	
\usepackage{amsfonts}
\usepackage{enumerate} 
\usepackage{colordvi} 
\usepackage{bm}
\usepackage{epsfig}
\usepackage{float}
\usepackage{verbatim}
\usepackage{subfig}
\usepackage[dvipsnames]{xcolor}
\usepackage{multirow,multicol}

\usepackage[colorinlistoftodos]{todonotes}
\usepackage{hyperref}
\usepackage{ulem}

\usepackage[colorinlistoftodos]{todonotes}
\usepackage{hyperref}

\newcommand{\bfk}{\mbox{\boldmath$k$}}

\usepackage{tikz}
\usetikzlibrary{shapes.geometric, arrows,positioning,shadows,trees,calc}
\tikzset{
  basic/.style  = {draw, text width=2cm, drop shadow, font=\sffamily, rectangle, minimum height = 0.7cm, minimum width = 0.5cm},
  root/.style   = {basic, rounded corners=2pt, thin, align=center,
                   fill=green!80, minimum width = 0.5cm},
  level 2/.style = {basic, rounded corners=6pt, thin,align=center, fill=cyan!10,
                     minimum width = 0.5cm},
  level 3/.style = {basic, thin, align=left, fill=cyan!80,  minimum width = 0.4cm}
}

\tikzstyle{eftofde} = [rectangle,  minimum width=0.5cm, minimum height=0.5cm, text centered, draw=black, fill=yellow!80]
\tikzstyle{evoleq} = [rectangle,  minimum width=0.5cm, minimum height=0.5cm, text centered, draw=black, fill=orange!90]
\tikzstyle{covth} =[rectangle, minimum width=0.5cm, minimum height=0.5cm, text centered, draw=black, fill=orange!50]

\tikzstyle{linear} = [rectangle, text centered, draw=black, fill=red!20]
\tikzstyle{1loop} = [rectangle, text centered, draw=black, fill=red!50]
\tikzstyle{1halo} = [rectangle,  text centered, draw=black, fill=red!80]

\tikzstyle{reaction} = [circle,  text centered, draw=black, fill=green!80]
\tikzstyle{arrow} = [thick,->,>=stealth]
\tikzstyle{darrow} = [dashed,->,>=stealth]

\tikzset{
diagonal fill/.style 2 args={fill=#2, path picture={
\fill[#1, sharp corners] (path picture bounding box.south west) -|
                         (path picture bounding box.north east) -- cycle;}},
reversed diagonal fill/.style 2 args={fill=#2, path picture={
\fill[#1, sharp corners] (path picture bounding box.north west) |- 
                         (path picture bounding box.south east) -- cycle;}}
}

\title[Model independent nonlinear reaction]{Fast and accurate predictions of the nonlinear matter power spectrum for general models of Dark Energy and Modified Gravity}

\author[B. Bose ]{B. Bose$^{1,2}$\thanks{E-mail:ben.bose@ed.ac.uk},
M. Tsedrik$^{1}$,
J. Kennedy$^{2,3}$,
L. Lombriser$^{2}$,
A. Pourtsidou$^{1,4}$, 
A. Taylor$^{1,4}$. 
\\
$^{1}$Institute for Astronomy, University of Edinburgh, Royal Observatory, Blackford Hill, Edinburgh, EH9 3HJ, U.K. \\ 
$^{2}$D\'epartement de Physique Th\'eorique, Universit\'e de Gen\`eve, 24 quai Ernest Ansermet, 1211 Gen\`eve 4, Switzerland. \\
$^{3}$The Global Academy of Agriculture and Food Systems, University of Edinburgh, Easter Bush Campus, Charnock Bradley Building, EH25 9RG.
\\
$^{4}${Higgs Centre for Theoretical Physics, School of Physics and Astronomy,
The University of Edinburgh, Edinburgh EH9 3FD, UK.}
}

\date{Accepted XXX. Received YYY; in original form ZZZ}

\pubyear{2022}

\begin{document}
\label{firstpage}
\pagerange{\pageref{firstpage}--\pageref{lastpage}}
\maketitle

\begin{abstract}
We embed linear and nonlinear parametrisations of beyond standard cosmological physics in the halo model reaction framework, providing a model-independent prescription for the nonlinear matter power spectrum. As an application, we focus on Horndeski theories, using the Effective Field Theory of Dark Energy (EFTofDE) to parameterise linear and quasi-nonlinear perturbations. In the nonlinear regime we investigate both a nonlinear parameterised-post Friedmannian (nPPF) approach as well as a physically motivated and approximate phenomenological model based on the error function (Erf). We compare the parameterised approaches' predictions of the nonlinear matter power spectrum to the exact solutions, as well as state-of-the-art emulators, in an evolving dark energy scenario and two well studied modified gravity models, finding sub-percent agreement in the reaction using the Erf model at $z\leq1$ and $k\leq 5~h/{\rm Mpc}$. This suggests only an additional 3 free constants, above the background and linear theory parameters, are sufficient to model nonlinear, non-standard cosmology in the matter power spectrum at scales down to $k \leq 3h~/{\rm Mpc}$ within $2\%$ accuracy. We implement the parametrisations into ver.2.0 of the {\tt ReACT} code: \href{https://github.com/nebblu/ACTio-ReACTio}{\tt ACTio et ReACTio}. 
\end{abstract}



\section{Introduction} \label{sec:introduction}

Fundamental models of nature generally begin with an action, which when combined with the principle of least action, gives us the temporal and spatial dynamics of the system. For the physical system that is our Universe (U), the action is widely accepted to be the action associated with general relativity (GR), the Einstein-Hilbert (EH) action, together with a matter contribution and cosmological constant
\begin{equation}
S_{U} = S_{\rm EH} + S_{\rm M} = \int d^{4}x \sqrt{-g} \left[ \frac{R}{2\kappa^2} -\frac{\Lambda}{\kappa^2}  \right] + S_{\rm M} \,,
\label{eq:eh} 
\end{equation}
where $\kappa^2 = 8 \pi G_{\rm N}$, $G_{\rm N}$ being Newton's gravitational constant and $R$ is the 4-dimensional Ricci scalar that gauges the curvature of spacetime. $S_{\rm M}$ is the action of the matter content of the Universe, usually approximated by a perfect, pressureless fluid, but in general will contain all Standard Model fields. $\Lambda$ is the (cosmological) constant that can appear naturally in a 4-dimensional action without violating preferred symmetries \citep[see, for example,][]{Fernandes:2022zrq}. This constant is measured to be non-zero by a suite of cosmological probes such as the cosmic microwave background (CMB) radiation \citep{Planck:2018vyg}, type 1a supernovae \citep{Riess:1998cb,Perlmutter:1998np}, and optical galaxy surveys \citep[see, for example,][]{eBOSS:2020yzd}. This has led to the standard model of cosmology, $\Lambda$CDM, where CDM stands for cold dark matter\footnote{CDM is the primary matter component in this model, outweighing baryonic matter five fold according to cosmological and astrophysical measurements such as the CMB.}.  

Consistently, we would expect a non-zero cosmological constant from quantum field theory (QFT) predictions, as all vacuum states of standard model particle fields will contribute an energy density, $\rho_{\rm vac}$, to the Universe that appears as a constant in the model's action. Unfortunately, this results in one of the biggest problems in physics \citep[see][for a review]{Martin:2012bt}. The first aspect of the problem is that our na\"{i}ve (QFT) predictions for the energy density of $\Lambda$ is at least 60 orders of magnitude larger than the (cosmological) measured value\footnote{This depends on the energy scales we are considering in the QFT calculation.}. We can still make a (fine) tuning of the `bare' constant $\Lambda_{\rm bare}$ in the potentials of these fields to cancel the other vacuum energy contributions to yield the observed value for $\Lambda$. 

One might be fine with this, after all QFTs are used to removing divergences through renormalisation techniques. The real problem is that we need to repeatedly fine tune every time a new energy scale or particle field is considered which changes $\rho_{\rm vac}$ (this can also happen through phase transitions) \citep[see][for an overview]{Padilla:2015aaa}. In other words, the value of $\Lambda$, which is a low energy physics parameter, is incredibly sensitive to the high energy physics, which is not technically natural and in apparent opposition to our wide spread employment of effective field theories. These two aspects of the problem are often referred to collectively as the `cosmological constant problem'. We refer the interested reader to the seminal paper by \cite{Weinberg:1988cp} for a review and the famous no-go theorem which implicitly delineates possible solutions to the problem.

Prospective solutions to these problems include gravitationally screening the vacuum energy from our observations by using a scalar field \citep[for example,][]{Charmousis:2011bf,Appleby:2020njl,SobralBlanco:2020too,Khan:2022bxs} or using extra spacetime dimensions \citep[for example,][]{Burgess:2004kd}. These solutions would of course also need to produce a small residual energy that can be used to explain our cosmological measurements, in particular those associated with an accelerated spatial expansion. This distinct issue can be called the `dark energy problem', which may be explained through a variation in the fundamental constants of nature such as Newton's gravitational constant, or having the acceleration driven by a scalar field \citep[see][for a general parameterisation of such options]{Thomas:2022phf}. 

The dark energy and cosmological constant problems motivate a minimal extension of \autoref{eq:eh} to include a single extra scalar degree of freedom, $\phi$, which is both physically and theoretically acceptable, i.e., not allowing for negative energies for example, and can encapsulate one or more cosmological constant problem solutions. Such an extension is found in the well studied Horndeski (H) scalar-tensor theory \citep{Horndeski:1974wa}. This is the most general, Lorentz-covariant scalar-tensor theory in 4 spacetime dimensions that yields second-order equations of motion, a basic condition for the physical viability of the theory, i.e., it is ghost-free. A universe described by Horndeski gravity is given as 
\begin{align}
S_{U} &= S_{\rm H} + S_{\rm M} = \int d^{4}x \sqrt{-g} \nonumber \\ 
& \big[ G_2(\phi,X) - G_3(\phi,X)\Box\phi +G_4(\phi,X)R  \nonumber \\
& + G_{4,X}(\phi,X)[(\Box \phi)^2-(\nabla_\mu \nabla_\nu \phi)^2]   \nonumber \\
&+ G_5(\phi,X)G_{\mu \nu}\nabla^\mu\nabla^\nu\phi  \nonumber \\
&- \frac{1}{6}G_{5,X}(\phi,X) [(\Box\phi)^3-3\Box\phi(\nabla_\mu\nabla_\nu\phi)^2+2(\nabla_\mu\nabla_\nu\phi)^3] \Big] \nonumber \\ & + S_{\rm M} \,,
\label{eq:horndeski}
\end{align}
where each $G_i(\phi,X)$, $i=2,3,4,5$ is a free function of the scalar field $\phi$ and its canonical kinetic term $X = -(\partial \phi)^2/2$, and $G_{i,X}(\phi,X) = \partial G_{i}/\partial X$. 

This opens up a very large theory space which needs to be trimmed down with observational data. We have very strong data constraints at small spatial scales, i.e., within the Solar System and at astrophysical scales \citep{1993tegp.book.....W,Will:2014kxa}, showing gravity is highly consistent with GR in this regime. We also have high quality observational data from cosmology, primarily from the CMB which is associated with early cosmological times. This allows new theoretical models most phenomenological freedom at large temporal and spatial scales as they must recover CMB and solar system observations. The small spatial scale constraints can be evaded using so called {\it screening mechanisms} \citep[see][for reviews]{Koyama:2018som,Burrage:2017qrf} that force predictions of modified gravity models back to those of GR locally, while early time measurements like the CMB can easily be recovered through appropriate time evolution of $\phi$. 

An obvious late time cosmological data set directly related to gravity is the large scale structure of the Universe (LSS). A key summary statistic of this is the two point correlation function or {\it power spectrum} (in Fourier space) of the cosmological matter field. A prime science goal then becomes the production of accurate predictions of the matter power spectrum in general theories beyond-$\Lambda$CDM. For the Horndeski class of models, this is a nontrivial task as there are an additional four free functions of space and time to contend with, beyond the matter content and metric freedoms. Of course, one can always choose particular forms for the $G_i(\phi,X)$ and then produce predictions for the 2-point correlations of matter. This approach allows one to fully specify how matter should cluster at all physical scales, and there are many tools and models that do just that to varying degrees of accuracy \citep{Schmidt:2009yj,Lombriser:2014dua,Arnold:2021xtm,Cataneo:2018cic,Bose:2020wch,Hernandez-Aguayo:2021kuh,Puchwein:2013lza,Brax:2013fna,Brax:2014yla,Joudaki:2020shz,Winther:2017jof}. 

If on the other hand we choose not to specify a particular model, we are required to parameterise both the linear and nonlinear scales i.e., the large and small physical scales of LSS respectively. At linear scales and for the Horndeski class of models, we can opt to perform a Taylor expansion of the $G_i$ functions and truncate at some order. Linear theory can then be characterised by a small number of free functions of time but with no unique specification in the nonlinear regime. This describes the approach of the Effective Field Theory of Dark Energy (EFTofDE) by \cite{Gubitosi:2012hu,Bloomfield:2012ff} \citep[also see][for a review]{Frusciante:2019xia}. Note that if we wish to be even more general than Horndeski we can directly parametrise the linear relation between matter and the gravitational potential. 

On nonlinear scales, a parameterisation framework one can consider is the nonlinear parameterised post-Friedmannian (nPPF), which captures modified gravity or dark energy effects \citep{Lombriser:2016zfz}. Both linear and nonlinear parameterisations then need to be consistently embedded in some more comprehensive predictive framework in order to be able to confront theory with  LSS observations. 

For past galaxy surveys the precision of the data did not call for high accuracy in the power spectrum modelling, \citep[as argued in][]{SpurioMancini:2019rxy,Traykova:2019oyx}. This changes with the next generation of surveys (Stage-IV) such as Euclid\footnote{\url{http://euclid-ec.org}}~\citep{Laureijs:2011gra} and the Vera C. Rubin Observatory’s Legacy Survey of Space and Time (LSST)\footnote{\url{https://www.lsst.org/}}~\citep{LSST:2008ijt}. These surveys will provide a significant reduction in statistical errors, errors which will be lowest in the nonlinear regime. With such precision, we have the opportunity to greatly constrain deviations to $\Lambda$CDM, including the well defined model space within \autoref{eq:horndeski}. This is contingent on whether or not we can accurately and efficiently map these deviations to the matter power spectrum. Typically, to remain unbiased in our constraints on Nature, $\mathcal{O}(1)\%$ is quoted as being the target accuracy for theoretical predictions \citep[see][for example]{Blanchard:2019oqi}. But this is not sufficient. We also require this map to be computationally efficient enough to perform data analyses. Without accuracy, we forfeit trust in our constraints. Without conciseness and efficiency we face major computational issues.

This paper provides a balance that satisfies these criteria. We mainly focus on the Horndeski class of models, embedding the EFTofDE and nPPF approaches into the halo model reaction framework \citep{Cataneo:2018cic,Giblin:2019iit,Cataneo:2019fjp,Bose:2019yjp,Bose:2020wch,Carrilho:2021rqo}, which is able to predict the nonlinear power spectrum for specified theories beyond-$\Lambda$CDM at $\mathcal{O}(1)\%$-level accuracy. We also present a completely model independent parametrisation of beyond-$\Lambda$CDM physics at nonlinear scales, which can be combined with similar parametrisations for the Universe's background expansion history and linear structure formation, giving a parametrisation for general deviations to $\Lambda$CDM. Thus, this work promotes the halo model reaction framework to being able to perform model independent predictions, a key step in the search for a more fundamental description of Nature in the cosmological, low energy regime.

This paper is organised as follows: in \autoref{sec:hmr} we begin at the observational end and look how to model the (halo model) {\it reaction}. In \autoref{sec:parameterisations} we jump to the theoretical end, and look how we can connect the ingredients of the reaction to an {\it action} of Nature, together with any additional degrees of freedom characterising nonlinear physics. In \autoref{sec:approximations} we give an overview of the mapping between the reaction and the parameterisations of gravity and dark energy, along with some key simplifying approximations one can consider. We also perform tests and provide motivations for these approximations. Then, in \autoref{sec:validation} we test the proposed parameterisations by comparing to exact predictions as well as emulators based on $N$-body simulations in an evolving dark energy scenario and two representative non-standard models of gravity. In \autoref{sec:summary} we summarise our results and conclude. In the appendices we provide full expressions for the linear and nonlinear parametrisations as well as illustrative examples and comparisons within specific non-standard models of gravity.


\section{Halo model reaction} \label{sec:hmr}
The leading order moment of the cosmological matter distribution is the {\it nonlinear matter power spectrum}, $P_{\rm NL}(k,z)$. This Fourier space quantity captures most of the matter clustering information at all scales \citep[see][for a review]{Bernardeau:2001qr}. Following the halo model \citep[see][for a review]{Cooray:2002dia} based approach of \cite{Cataneo:2018cic}, in a target theory of cosmology and gravity this quantity can be modelled as 
\begin{equation}
 P_{\rm NL}(k,z) = \mathcal{R}(k,z) P^{\rm pseudo}_{\rm NL}(k,z)\,,
 \label{eq:nonlinpk}
\end{equation}
where $P^{\rm pseudo}_{\rm NL}(k,z)$ is called the pseudo power spectrum. This is defined as the power spectrum of a $\Lambda$CDM universe but whose initial conditions have been set so as to match the target, beyond-$\Lambda$CDM, theory's linear total matter power spectrum $P_L(k,z)$ at some target redshift, $z$. The reason for making such a definition is that it guarantees the halo mass functions in the target and pseudo universes are similar since they will have the same linear clustering by definition. This results in a smoother transition between the clustering statistics in the inter- and intra-halo regimes. This quantity can be modelled in a number of ways, for example by using existing halo model based fitting functions such as \texttt{HMCode} \citep{Mead:2015yca,Mead:2016zqy,Mead:2020vgs} or for target theories that only predict a redshift dependent, but scale independent rescaling of the linear spectrum,  $\Lambda$CDM-based emulators such as \texttt{EuclidEmulator2} \citep{Knabenhans:2020gdo} or \texttt{bacco} \citep{2021MNRAS.507.5869A} can be used by tuning the spectrum amplitude parameter to match the modified cosmology's linear spectrum.

The function $\mathcal{R}(k,z)$ represents all the corrections to the pseudo spectrum coming from  nonlinear beyond-$\Lambda$CDM physics. Following \cite{Cataneo:2019fjp,Bose:2021mkz} we can write this as 
\begin{equation}
    \mathcal{R}(k)=\frac{\left(1-f_{\nu}\right)^{2} P_{\mathrm{hm}}^{(\mathrm{cb})}(k)+2 f_{\nu}\left(1-f_{\nu}\right) P_{\mathrm{hm}}^{(\mathrm{cb} \nu)}(k)+f_{\nu}^{2} P_{\mathrm{L}}^{(\nu)}(k)}{ P_{\rm hm}^{\rm pseudo}(k,z)} \, ,
    \label{eq:reaction}
\end{equation}
with the subscript
‘hm’ standing for halo model, $\rm (m) \equiv (cb+\nu) $, cb for CDM and baryons, $\nu$  for massive neutrinos and $f_\nu = \Omega_{\nu,0}/\Omega_{m,0}$ being the massive neutrino energy density fraction at $z=0$. The effects of massive neutrinos are included linearly through the weighted sum of the nonlinear cb halo model and linear massive neutrino spectra following the findings of \cite{Agarwal:2010mt}. We note that we do not consider massive neutrino effects in this work, but have included them in the expressions to highlight the generality of this approach \citep[see][for a study with massive neutrinos]{Bose:2021mkz}. 

The individual components are given by 
\begin{align}
    P_{\mathrm{hm}}^{(\mathrm{cb} \nu)}(k) \approx & \sqrt{P_{\mathrm{hm}}^{(\mathrm{cb})}(k) P_{\mathrm{L}}^{(\nu)}(k)} \, , \\
    P_{\mathrm{hm}}^{(\mathrm{cb})}(k) =& \left[(1-\mathcal{E}) e^{-k / k_{\star}}+\mathcal{E}\right]  P_{\mathrm{L}}^{(\mathrm{cb})}(k)+P_{\mathrm{1h}}^{(\mathrm{cb})}(k)  \, , \label{eq:1hcb}  \\  
  P_{\rm hm}^{\rm pseudo}(k,z) = &   P_{\rm L} (k,z) + P_{\rm 1h}^{\rm pseudo}(k,z), \label{Pk-halos} 
  \end{align}
  where the parameters are given by
  \begin{align} 
  \mathcal{E}(z) =&  \lim_{k\rightarrow 0} \frac{(1-f_\nu)^2 P_{\rm 1h}^{\rm (cb)}(k,z)}{ P_{\rm 1h}^{\rm pseudo}(k,z)} , \label{mathcale} \\ 
   k_{\rm \star}(z) = & - \bar{k} \left(\ln \left[ 
    \frac{T_1(\bar{k},z) \pm T_2(\bar{k},z)}{(1-f_\nu)^2 P^{\rm (cb)}_{\rm L}(\bar{k},z) (1-\mathcal{E}(z))  }  \right] \right)^{-1} \,. \label{kstar}
\end{align}
 We take the `+' root if $\mathcal{E}>1$, otherwise we take the `-' root. The $T_i$ terms are given by
\begin{align}
    T_1(k,z) =& f_\nu^2 P_{\mathrm{L}}^{(\nu)}(k,z) + P_{\rm hm}^{\rm pseudo}(k,z) \mathcal{R}_{\rm SPT}(k,z) \label{eq:T1} \nonumber \\ & - (1-f_\nu)^2[\mathcal{E}(z) P^{\rm (cb)}_{\rm L}(k,z) + P_{\mathrm{1h}}^{(\mathrm{cb})}(k,z)] \, , \\
     T_2(k,z) =& 2 \sqrt{f_\nu^2 P_{\rm hm}^{\rm pseudo}(k,z)  P_{\mathrm{L}}^{(\nu)}(k,z)  \mathcal{R}_{\rm SPT}(k,z)} \, , \label{eq:T2}
\end{align}
where $\mathcal{R}_{\rm SPT}(k,z)$ is the 1-loop standard perturbation theory (SPT) \citep{Bernardeau:2001qr} prediction for the reaction given by \autoref{eq:reaction}-\ref{Pk-halos} but with the replacements $P_L(k,z) \rightarrow P_{\rm 1-loop}(k,z)$ and $P_L(k,z)^{\rm (cb)} \rightarrow  P_{\rm 1-loop}(k,z)^{\rm (cb)}$ and $\mathcal{E} = 1$. As in \cite{Cataneo:2018cic} the default scale where we calculate $k_{\rm \star}$ is set to $\bar{k}=0.06~h~{\rm Mpc}^{-1}$. 

We see that \autoref{eq:reaction} depends on three general predictions for beyond-$\Lambda$CDM theories: the 2-halo term which we have approximated by the linear power spectrum $P_{\rm L}$, the quasi-nonlinear power spectrum given by the 1-loop perturbation theory power spectrum $P_{\rm 1-loop}$, and the highly nonlinear power spectrum given by the 1-halo term $P_{\rm 1h}$. The computation of these quantities requires the specification of the matter density fluctuations at different physical scales. The first two regimes (linear and quasi-nonlinear) are perturbatively derived up to 3rd order in the linear density fluctuation $\delta_L$,  while the fully nonlinear quantity, $\delta_{\rm NL}$, can be obtained using the assumptions of spherical collapse \citep{Cooray:2002dia}. Both these routes require us to solve differential equations representing energy and momentum conservation on a cosmological background. Our Universe's spacetime metric is well described by the  Friedman-Lemaître-Robertson-Walker (FLRW) metric, whose background expansion is described by the Hubble parameter $H(a) \equiv \dot{a}/a$, where $a$ is the scale factor and an over-dot represents a derivative with respect to the metric time $t$. 

Further, the conservation equations rely on the relation between the gravitational potential and the matter density fluctuation: the {\it Poisson equation}. In particular, we consider the Poisson equation in the perturbative limit,  only valid up to quasi-nonlinear scales, as well as the fully nonlinear expression, valid at all scales
\begin{align}
-\left(\frac{k}{a H(a)}\right)^2\Phi_{\rm QNL} (\bfk,a)=&
\frac{3 \Omega_{\rm m}(a)}{2} \mu(k,a) \,\delta_{\rm QNL}(\bfk,a) + S(\bfk,a) \, , \label{eq:poisson1} \\ 
-\left(\frac{k}{a H(a)}\right)^2\Phi_{\rm NL} (\bfk,a)=&
\frac{3 \Omega_{\rm m}(a)}{2} [1+ \mathcal{F}(k,a)] \,\delta_{\rm NL}(\bfk,a) \, ,
\label{eq:poisson2}
\end{align} 
where $\Omega_{\rm m}(a)=\Omega_{\rm{m},0} H_0^2/(H(a)^2a^3)$, $\Omega_{\rm{m},0}$ being the total matter fraction today. $\Phi$ is the gravitational potential in the time-time component of the perturbed FLRW metric. This can be identified with the Newtonian gravitational potential in the non-relativistic limit, valid for the curvatures and velocities we consider. The subscripts QNL and NL denote `quasi-nonlinear' and `nonlinear' respectively. One should further note that \autoref{eq:poisson1} and \autoref{eq:poisson2} also assume a spherically symmetric density distribution.  

The additional functions in \autoref{eq:poisson1} and \autoref{eq:poisson2} are as follows: $\mu(k,a)$ characterises the linear modification to GR, $\mathcal{F}(k,a)$ is the nonlinear modification and  $S(\bfk,a)$ is a source term capturing modifications at 2nd and 3rd order in the linear matter density perturbations. The source term is given by \citep{Bose:2016qun}
\begin{align}
 S(\bfk,a) & =
\int\frac{d^3\bfk_1d^3\bfk_2}{(2\pi)^3}\,
\delta_{\rm D}(\bfk-\bfk_{12}) \gamma_2(\bfk_1, \bfk_2,a)
\delta(\bfk_1)\,\delta(\bfk_2)
\nonumber\\
 & +
\int\frac{d^3\bfk_1d^3\bfk_2d^3\bfk_3}{(2\pi)^6}
 \delta_{\rm D}(\bfk-\bfk_{123})
\gamma_3(\bfk_1, \bfk_2, \bfk_3,a) \nonumber \\ & \quad \quad \times
\delta(\bfk_1)\,\delta(\bfk_2)\,\delta(\bfk_3) \, ,
\label{eq:Perturb3}
\end{align}
introducing two additional functions $\gamma_2$ \& $\gamma_3$ characterising quasi-nonlinear modifications to the Poisson equation \citep[see][for explicit expressions for these in the Horndeski class of models]{Bose:2016qun}. The functions $\gamma_2$, $\gamma_3$ and $\mathcal{F}$ all encode details regarding the potential screening mechanism of the theory under consideration. On this point, it is worth noting that  for general theories beyond-$\Lambda$CDM such mechanisms may not be present, in which case the spherical density distribution approximation  assumed in  \autoref{eq:poisson1} \& \autoref{eq:poisson2} may break down \citep{Thomas:2020duj}. For the modified gravity models considered in this work, which have some method of screening, this appears to be a reasonable approximation \citep{Noller:2013wca}. For a study of screened and unscreened models in the Horndeski class see \cite{Noller:2020lav}. 

In total, the halo model reaction, and so the nonlinear power spectrum, requires specification of four functions of space and time - one for the background $H(a)$, one for the linear regime $\mu(k,a)$, two for the quasi-nonlinear regime $\gamma_2(\bfk_1,\bfk_2,a)$ \& $\gamma_3(\bfk_1,\bfk_2,\bfk_3,a)$ and finally one for the fully nonlinear regime $\mathcal{F}(k,a)$. In principle these functions are not completely independent, and one should have $\mathcal{F} \rightarrow \mu$ in the linear limit. We investigate the importance of respecting this limit in \autoref{sec:validation}. Finally, we remind the reader that all these functions are required to compute the key ingredients of $\mathcal{R}$ (and hence $P_{\rm NL}$): $P_{\rm L}$, $P_{\rm 1-loop}$ and $P_{\rm 1h}$. 

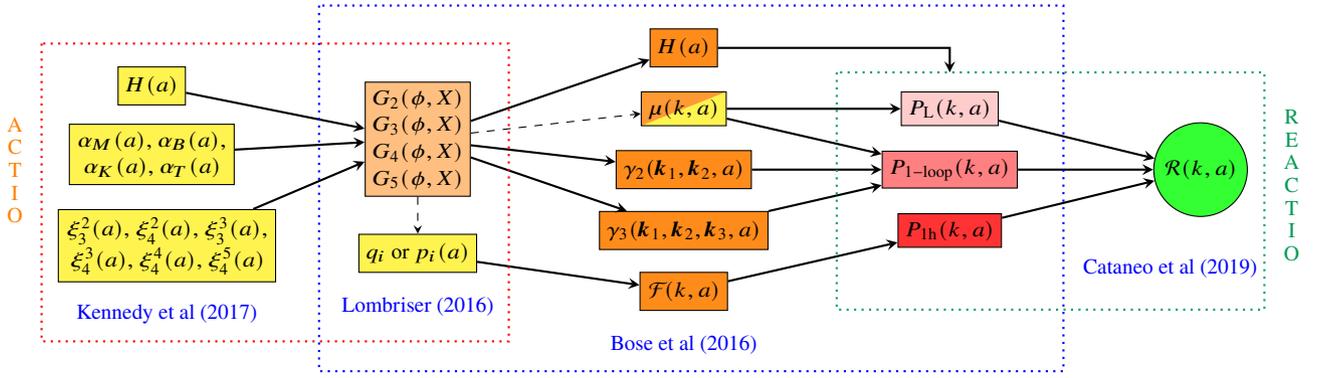
\begin{figure*}
\centering
\begin{tikzpicture}[node distance=2cm]
\node (background) [eftofde, xshift=-10cm, align=center] {$H(a)$};
\node (1st) [eftofde , below of=background,yshift = 1.1cm, align=center] {$\alpha_M(a),\alpha_B(a),$ \\ $\alpha_K(a),\alpha_T(a)$};
\node (2nd) [eftofde , below of=1st,yshift = 0.8cm, xshift=0.2cm, align=center] {$\xi_3^2(a), \xi_4^2(a), \xi_3^3(a),$ \\  $\xi_4^3(a), \xi_4^4(a), \xi_4^5(a)$};

\node (covth) [covth , right of=1st, xshift=1.5cm, yshift = 0.2cm, align=center] {$G_2(\phi,X)$ \\ $G_3(\phi,X) $\\$G_4(\phi,X)$\\ $G_5(\phi,X)$};

\node (scol) [eftofde , below of=covth, yshift = 0.5cm, align=center] {$q_i$ or $p_i(a)$};

\node (hubble) [evoleq ,right of=covth, xshift=1.5cm,yshift=1.2cm] {$H(a)$};

\node(mu)[diagonal fill={yellow!80}{orange!80},
      below of=hubble,yshift = 1.2cm,
      text centered, draw]{$\mu(k,a)$};

\node (gamma2) [evoleq , below of=mu,yshift = 1.2cm] {$\gamma_2(\bfk_1,\bfk_2,a)$};
\node (gamma3) [evoleq , below of=gamma2,yshift = 1.2cm] {$\gamma_3(\bfk_1,\bfk_2,\bfk_3,a)$};

\node (geff) [evoleq , below of=hubble,yshift = -1.2cm] {$\mathcal{F}(k,a)$};

\draw [arrow] (background) -- (covth);
\draw [arrow] (1st) -- (covth);
\draw [arrow] (2nd) -- (covth);

\draw [arrow] (covth) -- (hubble);
\draw [darrow] (covth) -- (mu);
\draw [arrow] (covth) -- (gamma2);
\draw [arrow] (covth) -- (gamma3);

\draw [arrow] (scol) -- (geff);

\draw [darrow] (covth) -- (scol);

\node (linear) [linear ,right of = mu,align=center, yshift=0cm,xshift=1.5cm] {$P_{\rm L}(k,a)$};

\node (1loop) [1loop ,below of = linear,align=center, yshift=1.2cm] {$P_{\rm 1-loop}(k,a)$};

\draw [arrow] (mu) -- (linear);

\draw [arrow] (mu) -- (1loop);
\draw [arrow] (gamma2) -- (1loop);
\draw [arrow] (gamma3) -- (1loop);

\node (1halo) [1halo ,below of = 1loop,align=center, yshift=1.2cm] {$P_{\rm 1h}(k,a)$};

\draw [arrow] (geff) -- (1halo);

\node (reaction) [reaction ,right of = 1loop,align=center, xshift=1.3cm] {$\mathcal{R}(k,a)$};

\draw [arrow] (linear) -- (reaction);
\draw [arrow] (1loop) -- (reaction);
\draw [arrow] (1halo) -- (reaction);

\draw[arrow,black,thick] ($(hubble.east)+(0,0)$)  -- (0.5,0.5) -- ($(linear.north)+(0.,0.25)$);

\draw[red,thick,dotted] ($(background.north west)+(-1,0.3)$)  rectangle ($(covth.south east)+(0.5,-1.9)$);

\node (klt) [below of=2nd, yshift=1.1cm, align=center] {\href{https://arxiv.org/pdf/1705.09290.pdf}{Kennedy et al (2017)}};

\draw[blue,thick,dotted] ($(covth.north west)+(-0.6,1)$)  rectangle ($(1loop.south east)+(0.6,-2.4)$);
\node (lomb) [below of=scol, align=center,yshift=1.3cm] {\href{https://arxiv.org/pdf/1608.00522.pdf}{Lombriser (2016)}};

\node (bk) [below of=gamma3, xshift = 0.cm, yshift = 0.5cm, align=center] {\href{https://arxiv.org/pdf/1606.02520.pdf}{Bose et al (2016)}};

\draw[ForestGreen!90,thick,dotted] ($(1halo.north west)+(-0.8,1.85)$)  rectangle ($(reaction.south east)+(0.4,-1.4)$);

\node (cat) [below of=reaction,  yshift = 0.7cm, xshift=-0.4cm, align=center] {\href{https://arxiv.org/abs/1812.05594}{Cataneo et al (2019)}};


\node (a) [left of=2nd, yshift=1.6cm, align=center, color=orange] {A};
\node (c) [left of=2nd, yshift=1.3cm, align=center, color=orange] {C};
\node (t) [left of=2nd, yshift=1.cm, align=center, color=orange] {T};
\node (i) [left of=2nd, yshift=0.7cm, align=center, color=orange] {I};
\node (o) [left of=2nd, yshift=0.4cm, align=center, color=orange] {O};

\node (r) [right of=reaction,  yshift = 0.7cm, xshift = -0.8cm, align=center, color=ForestGreen] {R};
\node (e) [right of=reaction,  yshift = 0.4cm, xshift = -0.8cm, align=center, color=ForestGreen] {E};
\node (a) [right of=reaction,  yshift = 0.1cm, xshift = -0.8cm, align=center, color=ForestGreen] {A};
\node (c) [right of=reaction,  yshift = -0.2cm, xshift = -0.8cm, align=center, color=ForestGreen] {C};
\node (t) [right of=reaction,  yshift = -0.5cm, xshift = -0.8cm, align=center, color=ForestGreen] {T};
\node (i) [right of=reaction,  yshift = -0.8cm, xshift = -0.8cm, align=center, color=ForestGreen] {I};
\node (o) [right of=reaction,  yshift = -1.1cm, xshift = -0.8cm, align=center, color=ForestGreen] {O};

\end{tikzpicture}
\caption{A rough schematic of the map from the Horndeski action in the EFTofDE parametrisation ($H,\alpha_i,\xi_i^j$) and nonlinear parameterisations ($p_i$ or $q_i$) to $\mathcal{R}(k,t)$. The yellow rectangles indicate the input functions of time (here parametrised by the scale factor $a$) or constants. The orange rectangles indicate the modifications to the Poisson equation. $\mu(k,a)$ is bi-coloured indicating we may choose to parametrise it directly instead of starting at the action level. We provide the main sources in the literature for each piece of the map along with a dotted box roughly indicating their associated piece. Note that the solid arrows can only reconstruct $G_i$ to the linear and quasi-nonlinear levels, which can in turn inform choices for $p_i$. The $p_i$ provide the nonlinear complement in the $G_i$.}
\label{fig:pipeline}
\end{figure*}

The right half of \autoref{fig:pipeline}  summarises the map from background and Poisson equations to the halo model reaction as described in this section. The left half of the figure will be the focus of the next section.  


\section{Parametrisations} \label{sec:parameterisations}
We now move away from the observational end and return to the starting point, the fundamental action of Nature. In particular, here we mostly focus on the Horndeski action given in \autoref{eq:horndeski}, but the approach can be trivially extended to further generality. 

As pointed out, given a specific form of the $G_i$ functions, the explicit functional forms of $H$, $\mu$, $\gamma_2$, $\gamma_3$ and $\mathcal{F}$ can be directly derived. But rather than  specifying the full covariant theory, i.e., 4 free functions of space and time, we ultimately wish to parameterise the action's predictions for cosmological matter clustering in terms of a few free constants. 

To do this, we split LSS into three regimes: the {\it background} \& {\it linear}, {\it quasi-nonlinear}  and the {\it nonlinear}. The background, linear and quasi-nonlinear regimes will follow the well studied EFTofDE program \citep{Gubitosi:2012hu,Bloomfield:2012ff}. For the nonlinear regime we will consider two different parameterisations. One is the established nonlinear parameterised post-Friedmannian (nPPF) approach \citep{Lombriser:2016zfz}. The other parametrisation we propose here is phenomenological and is based on some well known screening mechanisms. We begin by parameterising the  background \& linear regime. 


\subsection{Background \& Linear: Effective field theory of dark energy} \label{sec:eftofde}

Among the methods to generically parameterise beyond-$\Lambda$CDM physics on cosmological scales, the methods of Effective Field Theory (EFT) have proven to be particularly useful. It is simply necessary to determine which symmetries one wishes the action to have before constructing various operators out of the fields and derivatives of the fields. One can trust the predictions made with an EFT as long as it is made at an energy scale below the `cutoff' of the theory, beyond which the validity of the EFT breaks down.

While not being an EFT in this strict sense, the EFTofDE is constructed in a similar manner and is capable of describing the dynamics of the cosmological background and perturbations in Horndeski theory in a generic manner. The EFTofDE approach breaks time diffeomorphism invariance of the cosmological background by choosing a particular gauge. By doing this one is able to form a theory out of operators which only respect spatial diffeomorphism invariance.

In constructing the EFTofDE action one begins by foliating spacetime with constant-time hypersurfaces. Utilising the complete freedom one has in choosing the coordinates of the theory we can set the scalar field to be only a function of time such that $\phi(x,t)\rightarrow \phi(t)$. In particular, we can choose 
\begin{equation}
\phi = t/ \kappa^2 \,.
\end{equation}
This choice is called the {\it unitary gauge} and in this gauge the scalar field perturbations vanish, being absorbed into the time-time component of the metric. %
The operators in the EFTofDE are the cosmological perturbations themselves. 
In the unitary gauge we are free to include operators in the EFT which are only spatially diffeomorphism invariant, such as $g^{00}$.

Let us denote the normal vector to each spatial hypersurface as
\begin{equation}
    n_{\mu} = - \frac{\partial_{\mu} \phi}{\sqrt{-(\partial \phi)^{2}}} \, .
\end{equation}
The induced spatial metric of each hypersurface is then given by $h_{\mu\nu} = g_{\mu\nu}+n_{\mu}n_{\nu}$. This allows us to include the extrinsic curvature which is given by the projection of the derivative of the normal vector along the the hypersurface, onto the hypersurface $K_{\mu\nu} = h_{\mu \sigma} \nabla^{\sigma} n_{\nu}$. With the induced metric, one can also compute the intrinsic curvature of each hypersurface given by the three-dimensional Ricci scalar $R^{(3)}$. 

Collecting relevant combinations of the invariants under residual spatial diffeomorphism symmetry gives the EFTofDE action, which is capable of describing the dynamics of the background and linear perturbations of Horndeski theory. The action is given by \citep[see, for example,][]{Kennedy:2017sof}
\begin{align}
S_{U,L} = & \: S^{(0,1)}+S^{(2)}+S_{M}[g_{\mu\nu},\Psi_m] \,,
\label{eftlag}\\
S^{(0,1)} = & \:\int d^{4}x \sqrt{-g}   \left[ \frac{\Omega(t)}{2 \kappa^2} R - \Lambda(t) - c(t) \delta g^{00} \right] \,,
\label{eq:s01} \\
S^{(2)} = & \int d^{4}x \sqrt{-g}  \Big[ \frac{M^{4}_2(t)}{2}(\delta g^{00})^2-\frac{\bar{M}^{3}_{1}(t)}{2} \delta K \delta g^{00}    \nonumber \\
& -\bar{M}_2^2(t)\Big( \delta K^2 -  \delta K^{\mu \nu}  \delta K_{\mu \nu} - \frac{1}{2} \delta R^{(3)}\delta g^{00} \Big) \Big] \,.
\label{eq:s2}
\end{align}
where $S_{U,L}$ represents the action of a Horndeski-universe that describes field dynamics up to the linear level in the matter and velocity perturbations. The $(0,1,2)$ represent the order in the perturbed quantities. 

In front of each term we include a free function of time called an EFT coefficient, giving a total of six free functions, $\{\Omega(t), \Lambda(t), c(t), M_2^4(t), \bar{M}_1^3(t), \bar{M}_2^2(t) \}$. Once we specify a metric, we also introduce any metric degrees of freedom. For FLRW this is the scale factor $a$, or equivalently the Hubble parameter $H(a)$. We can then employ the field equation constraints, which in the FLRW are the Friedmann equations
\begin{align}
    0 = & \kappa^2(2 c - \Lambda + \rho_\mathrm{m}) - 3 H^2(\Omega + a \Omega')  \, , \\ 
    0 = & \kappa^2 \Lambda + H[a H' ( 2 \Omega + a \Omega') + H ( 3\Omega + 3 a \Omega' + a^2 \Omega'')] \, ,
\end{align}
where we have dropped the time dependence in constituent parameters for compactness, and use the scale factor to parameterise time. A prime denotes a scale factor derivative and $\rho_\mathrm{m} $ is the matter density at $a$. The Friedmann equations reduce the number of free functions describing the background and linear perturbations to five. Solving these equations yields
\begin{align}
   c(a) &= -\frac{\rho_\mathrm{m}}{2} - \frac{a H [ H' ( 2 \Omega +a \Omega') + a H\Omega'']}{\kappa^2} \, , \label{eq:cmbasis} \\
   \Lambda(a)  =& -\frac{H [ a H'(2\Omega + a \Omega') + H(3\Omega + 3a\Omega'+a^2\Omega'')]}{\kappa^2}\, .
\end{align}
This means the free functions of the scale factor defining the background and linear theory would be $\{\Omega, H, M_2^4, \bar{M}_1^3,\bar{M}_2^2 \}$, which we will refer to as the {\it $M$-basis}. We can alternatively write the Hubble function as the solution to
\begin{equation}
     H(a): \, \, 0 =  (2 \Omega + a \Omega') H' +  a H \Omega '' + \frac{\kappa \rho_\mathrm{m}}{a H} + \frac{2 \kappa c}{a H}    \, , \label{eq:hubble}
\end{equation}
if we wish to specify $c$ instead of $H$ for example. 

Common in the literature is the {\it $\alpha$-basis} $\{ H , \alpha_M, \alpha_B, \alpha_K, \alpha_T \}$  which has a clearer physical interpretation of the effects of each function \citep[see, for example,][]{Bellini:2014fua}. We provide the map between the $\alpha$- and $M$-bases\footnote{Note the factor of `$-1/2$' difference in $\alpha_B$ between our expression and that of {\tt EFTCAMB} \citep{Frusciante:2016xoj} or \cite{Kennedy:2018gtx}, for instance.}
\begin{align} 
\alpha_M &= \frac{a (M^2)'}{M^2}  \, , \label{eq:am} \\ 
\alpha_B &= -\frac{a H \Omega'  + \kappa^2 \bar{M}^3_1  }{H M^2 \kappa^2} \, , \label{eq:mb}  \\
\alpha_K &= \frac{2c + 4 \bar{M}_2^4}{M^2 H^2} \, , \label{eq:ak} \\
\alpha_T &= -\frac{ \bar{M}_2^2}{M^2} \, , \label{eq:at}  
\end{align} 
where  $M^2 =\Omega~ \kappa^{-2} +\bar{M}_2^2$. Note that one can alternatively specify $M^2$ and solve for $H$. 

To end this section, another basis worth considering is the basis introduced in \cite{Kennedy:2018gtx}: $\{H, M^2, c_s^2, \alpha, \alpha_{B0} \} $ \citep[also see][]{Lombriser:2018olq}, which implicitly assumes $\alpha_T=0$ (see \autoref{sec:constraints} for motivation). This basis allows for some simple priors on the functions that ensure the theory has no ghost or gradient instabilities, i.e., negative energies or imaginary sound speeds. We will refer to this basis as the  $s$-basis. The priors to ensure stability on these functions are then simply $M^2,c_s^2,\alpha >0$, and $\alpha_{B0}$ is constant\footnote{We note that this basis does not ensure the absence of a tachyonic instability \citep{Gsponer:2021obj}.}. The map between the $s$- and $\alpha$-bases is given by
\begin{align} 
 c_s^2 = & \frac{2}{\alpha} \Big[\frac{a \alpha_B'}{2} - (1+\alpha_T) \left(1-\frac{\alpha_B}{2}\right)^2 \nonumber \\ 
 & + \left( 1+ \alpha_M -\frac{a H'}{H} \right) \left(1-\frac{\alpha_B}{2}\right) - \frac{\rho_\mathrm{m}}{2 H^2 M^2} \Big] \, , \label{eq:cs2}  \\ 
\alpha =& \alpha_K + \frac{3}{2} \alpha_B^2 \, ,  \label{eq:alpha}
\end{align} 
where $c_s$ is the speed of sound, while $\alpha_{B0} = \alpha_B(a=1)$ is the boundary condition ($\alpha_B$'s value today) specified to solve the differential equation given by \autoref{eq:cs2}. 

\bigbreak

In what follows we will stick with the $\alpha$-basis and implement this as the default basis in the accompanying code \href{https://github.com/nebblu/ACTio-ReACTio}{\tt ACTio et ReACTio}. We provide the explicit form of the linear modification to the Poisson equation in this basis in \autoref{app:alphamu}. We leave it to the user to perform the transformation from their preferred basis to the $\alpha$-basis, and provide an accompanying notebook \href{https://github.com/nebblu/ACTio-ReACTio/tree/master/notebooks}{{\tt GtoPT.nb}} that performs some of these transformations.


\subsection{Quasi-nonlinear: Covariant theory map} \label{sec:eftofdeqsa}

To fully specify the halo model reaction, we need to go beyond the linear matter perturbations. In particular, we also require the 2nd and 3rd order density perturbations to solve for the 1-loop power spectrum entering $\mathcal{R}_{\rm SPT}$ in \autoref{eq:T1}-\ref{eq:T2}.  This requires us to expand to fourth order in the metric perturbation $\delta g^{00}$ and extrinsic curvature $\delta K^{\mu \nu}$ in \autoref{eftlag}. This has been done in \cite{Cusin:2017mzw} and has been used to calculate the 1-loop spectrum in \cite{Cusin:2017wjg}.  Further, in \cite{Kennedy:2017sof} the authors relate the EFTofDE functions up to a given order to the corresponding covariant theory's Lagrangian $G_i$ functions as 
\begin{equation}
    G_i(\phi,X) = g_i(\phi,X) + \Delta G_i(\phi,X) \, , \label{eq:gieft}
\end{equation}
where $g_i$, $i \in \{2,3,4,5 \}$, are well-defined functions of $\phi$, $X$ and the lower order EFTofDE parameters, e.g., $\{ H , \alpha_M, \alpha_B, \alpha_K, \alpha_T \}$. The other terms are given as 
\begin{align}
    \Delta G_{2,3} = &  \sum_{n>2} \xi_n^{(2,3)}(\phi) \Big( 1+ X\kappa^4 \Big)^n \, , \\ 
        \Delta G_{4,5} = &  \sum_{n>3} \xi_n^{(4,5)}(\phi) \Big( 1+ X\kappa^4 \Big)^n \, , 
\end{align}
where $\Delta G_i$ are higher order corrections to the covariant action and $\xi_n^i(\phi)$ are higher order EFTofDE functions, $X$ again being the scalar field canonical kinetic energy term.

A particular covariant theory is specified once $\xi_n^i$ are given for all $n\in \mathbb{N}$, but if we truncate at some order $n_{t}$, we specify the subset of Horndeski theories which are identical on scales described by the EFTofDE up to $ \xi_{n_t}^{i}$. Up to 3rd order in the matter density perturbation, we introduce 6 new functions with $n_t = 4$. Together with the background and linear order functions, this gives a total of 11 free functions of time for the quasi-nonlinear scales. The $G_i$ given in \autoref{eq:gieft} can then be related to $\mu$, $\gamma_2$ and $\gamma_3$ by the map provided in the Appendices of \cite{Bose:2016qun,Takushima:2015iha}. 

\bigbreak 

In \autoref{sec:efttopoisson} we provide the map between the 5 linear EFTofDE functions in the $\alpha$-basis and the linear modification to the Poisson equation, $\mu$, used in \autoref{eq:poisson1}. The 2nd and 3rd order functions $\gamma_2$ and $\gamma_3$ (see \autoref{eq:Perturb3}) are significantly more complicated but can be derived by using the map from the EFTofDE to $G_i(\phi,X)$ provided in \cite{Kennedy:2017sof} and then the $G_i(\phi,X)$ to $\gamma_2$ \& $\gamma_3$ given in \cite{Bose:2016qun}. The map, although not reproduced here in full, is given in detail in a {\tt Mathematica} notebook provided in the \href{https://github.com/nebblu/ACTio-ReACTio}{\tt ACTio et ReACTio} repository, \href{https://github.com/nebblu/ACTio-ReACTio/tree/master/notebooks}{{\tt GtoPT.nb}}. This being said, in \autoref{sec:approximations} we give support for the omission of $\gamma_2$ and $\gamma_3$ in the calculation of $\mathcal{R}$ for moderate to low modifications to gravity, and given the additional degrees of freedom we will introduce in the nonlinear regime. 

Having specified a route between the Horndeski action of nature and the linear and 1-loop power spectra, $P_{\rm L}(k,a)$ \& $P_{\rm 1-loop}(k,a)$, we now look at two methods of parameterising clustering in the highly nonlinear regime, characterised by the 1-halo term, $P_{\rm 1h}(k,a)$. This will then specify a full parameterisation of the halo model reaction $\mathcal{R}$, and consequently the nonlinear power spectrum, $P_{\rm NL}(k,a)$.


\subsection{Nonlinear} \label{sec:nonlinpar}

The effects of modified gravity on the nonlinear cosmic structure formation are captured by the effective deviation $\mathcal{F}$ from the gravitational constant in the nonlinear Poisson equation given in \autoref{eq:poisson2} and the cosmological background evolution. Specifically, the modified Poisson equation alters the evolution equation for the halo top-hat radius $R_{\rm TH}$ \citep[see, for example,][]{Schmidt:2008tn}. This quantity gives an estimate for $\delta_{\rm NL}$, needed to compute the 1-halo power spectrum. Here we discuss two parameterisations of $\mathcal{F}$.

\subsubsection{Nonlinear parametrised post-Friedmannian framework} \label{sec:pffF}

Following the nPPF approach of \citet{Lombriser:2016zfz}, the effective gravitational coupling for generic screening mechanisms and other suppression effects can be decomposed as a function of scale $r$
\begin{equation}
 1+\mathcal{F}(a,r) = A + \sum_i^{N_0} B_i \prod_j^{N_i} \mathcal{F}_{ij} \,, \label{eq:nlparam}
\end{equation}
where $\mathcal{F}_{ij}$ are some transition functions encapsulating screening or
other suppression effects such as a Yukawa suppression.
$N_0$ and $N_i$ characterise their respective number.
In the fully screened limit, the effective coupling reduces to $A$, typically unity, whereas it becomes $B_i$ in the fully unscreened limit, matching linear theory.
To parameterise these transitions, \citet{Lombriser:2016zfz} adopted a generalised form of the Vainshtein screening effect in the DGP braneworld model \citep{Dvali:2000hr}
\begin{equation}
 \mathcal{F} \sim b \left(\frac{r}{r_{\rm scr}}\right)^{a_f} \left\{ \left[1+\left(\frac{r_{\rm scr}}{r}\right)^{a_f}\right]^{1/b} - 1 \right\} \,, \label{eq:transition}
\end{equation}
where $r_{\rm scr}$ denotes the screening scale, which in general can be time, mass, and environment dependent.
The parameter $a_f$ (not to be confused with the scale factor) determines the radial dependence of the coupling in the screening limit along with $b$ that characterises an interpolation rate between the screened and unscreened limits.

Screening effects such as the chameleon \citep{Li:2011qda,Khoury:2003aq,Lombriser:2013eza}
symmmetron \citep{Hinterbichler:2010es, Taddei:2013bsk}, k-mouflage \citep{Babichev:2009ee,  Brax:2014yla}, and Vainshtein \citep{Vainshtein:1972sx, Schmidt:2009yj, Dvali:2000hr} mechanisms as well as other suppression effects such as the linear shielding mechanism \citep{Lombriser:2014ira} or Yukawa suppression, can be analytically mapped onto \autoref{eq:transition} by matching expressions in the limits of large and small $r$ and $r\rightarrow r_{\rm scr}$. The relevant expressions may be found in \cite{Lombriser:2016zfz}.
It is worth highlighting that the parameters of \autoref{eq:transition} for a given screening model may in principle be directly read off from \autoref{eq:horndeski} by employing the scaling method of \cite{McManus:2016kxu} \citep[also see][]{Renevey:2020tvr} and counting the powers of second and first spatial derivatives and the scalar field potential.
Note that the parameter $b$ may be understood as the choice of transition template used to approximately cast the screening effect into.
Alternatively to \autoref{eq:transition}, one could also adopt other transition functions such as a hyperbolic tangent, a sigmoid or an error function as we will propose in \autoref{sec:phenF}. For DGP, the choice of \autoref{eq:transition} with $b=2$ becomes exact.

To implement \autoref{eq:transition} in the spherical collapse model, one replaces $r/r_{\rm scr} \rightarrow y/y_{\rm scr}$, where $y$ is the normalised top-hat radius (\autoref{eq:ydef}). 
A single general element $N_0=N_1=1$ can then be described by seven parameters (or functions) $p_{1-7}$ in addition to $p_0=A$ (typically $=1$). The first three, $p_{1-3}$, determine $a_f$, $b$, and $B$. The other four are used to generally capture possible time, mass, and environmental dependencies of the dimensionless screening scale, which can be modelled as \citep{Lombriser:2016zfz}
\begin{equation}
 y_{\rm scr} = p_4 a^{p_5} \left(2G_N\,H_0 M_{\rm vir}\right)^{p_6} \left(\frac{y_{\rm env}}{y_{\rm h}}\right)^{p_7} \,,
 \label{eq:screeningscale}
\end{equation}
where $y_{\rm h}$ and $y_{\rm env}$ refer to the normalised radii of the the halo and the environment respectively, $H_0$ is the Hubble constant and $M_{\rm vir}$ is the virial mass of the halo\footnote{Note that in {\tt ReACT} we use the initial comoving top-hat radius, $R_{\rm th}$ (see \autoref{app:notation}), as an input parameter instead of mass, related as $M_\mathrm{vir} =4 \pi \bar{\rho}_{\mathrm{m}, i} (1+\delta_i) (a_i R_{\rm th})^3/3 \approx 4 \pi \Omega_
\mathrm{m,0} \rho_\mathrm{crit} R_{\rm th}^3/3$ with the critical density $\rho_{\rm crit}$ and $1+\delta_i \approx 1$.}. In this way, we can simplify \autoref{eq:transition} to \citep{Lombriser:2016zfz}
\begin{equation}
    \mathcal{F}_{\rm nPPF} = p_1 p_2 \frac{(1+s^{a_f})^{\frac{1}{p_1}}-1}{s^{a_f}} \, ,  \label{eq:nPPFF0}
\end{equation}
where 
\begin{equation}
    a_f = \frac{p_1}{p_1-1} p_3 \,  
\end{equation}
and $s = y_{\rm scr}/y_{\rm h}$. Note we have set $p_0=1$. The parameters $p_{1-7}$ can be computed from theory and in many cases take on trivial values (see \autoref{app:nppfforms}). It is worth highlighting here that the nPPF formalism has also been implemented in $N$-body simulations and cast into Fourier space~\citep{Hassani:2020rxd}, where it was shown to accurately match simulations of exact model implementations.

Finally, we consider the large, linear scale limit of $\mathcal{F}$. \autoref{eq:nPPFF} provides a parametrised function for the screening regime, where we have a transition to GR from some large scale modification. In this form, it does not capture any additional effects coming from say Yukawa suppression, typical of chameleon theories. Such phenomena may become relevant for the spherical collapse calculation at early times or for very large halo masses. In order to correctly capture this, we could either model the Yukawa suppression as another transition cast into \autoref{eq:nPPFF0} or simply augment \autoref{eq:nPPFF0} with the linear modification $\mu(k,a)$ as 
\begin{equation}
    \mathcal{F}_{\rm nPPF} = p_1 p_2 \frac{\left(1+s^{a_f}\right)^{\frac{1}{p_1}}-1}{s^{a_f}} \times (1-\mu(\hat{k},a)) \, . \label{eq:nPPFF}
\end{equation}
In this case we also need to perform the Fourier transform of $\mu(\hat{k},a)$, which is non-trivial. As a first order approximation, we parametrise this with a simple scaling of the inverse of the comoving initial top-hat radius $R_{\rm th}$ as 
\begin{equation}
    \hat{k} = \frac{10^{p_8}}{a^2 y_{\rm h} R_{\rm th}} \, , 
\end{equation}
where the dimensionless constant $p_8$ calibrates the Yukawa suppression. The Fourier transform can be made more sophisticated \citep[see, for example,][]{Hassani:2020rxd} but in \autoref{sec:validation} we find the impact of Yukawa suppression is negligible for the $f(R)$ models we consider, and so only include this augmentation for completeness. 
Further, \autoref{eq:nPPFF} would only be meaningful for a non-trivial scale dependent $\mu(k,a)$. For scale-independent theories one can absorb the scaling provided by $\mu(a)$ in the $p_2(a)$ parameter of \autoref{eq:nPPFF0}.

\subsubsection{Phenomenological parameterisation}\label{sec:phenF}

With its full freedom, the nPPF parameterisation is a very flexible way of modelling the nonlinear scales. It is able to capture various specific covariant theories exactly or to high accuracy (see \autoref{app:nppfforms} and \autoref{sec:validation}), and given a covariant theory, say from the Horndeski class, we can map its nonlinear Poisson modification to the $p_i$ parameters. On the other hand, if we remain agnostic about the covariant theory, 8 additional parameters, some of which may also be time dependent, poses computational issues as well as degrades the amount of cosmological and gravitational information we can extract due to degeneracies between these nuisance and the physical parameters of interest. 

With this in mind, we propose the following general and reduced parameterisation of $\mathcal{F}$ based on the error function (Erf). We have found this mimics the general profile of the effective gravitational constant in various modified gravity theories. Essentially we wish to capture a basic transition from unscreened to screened regimes. The simple form we adopt is given by 
\begin{equation}
    \mathcal{F}_{\rm Erf} = {\rm Erf}[ a y_{\rm h} 10^{\bar{J}}] \times  (1-\mu(\hat{k},a)) \, ,  \label{eq:ErfF}
\end{equation}
where as in the nPPF case, we use
\begin{equation}
    \hat{k} = \frac{10^{q_4}}{a^2 y_{\rm h} R_{\rm th}} \, , 
\end{equation}
and 
\begin{equation}
   \bar{J}  = q_1 - q_2 \log(R_{\rm th}) + q_3 \log (a y_{\rm env}) \, . \label{eq:Erfexp}
\end{equation}
$\mu$ is the linear modification to gravity. In the EFTofDE parameterisation $\mu$ is given in \autoref{eq:mueft}, but this can also be parametrised more generally \citep[see, for example,][]{Silvestri:2013ne,Kennedy:2018gtx,Srinivasan:2021gib}. 

The Erf model introduces 4 free constants:
\begin{itemize}
    \item[$\mathbf{q_1}$:]
    This parametrises the screening scale and goes as its inverse.
    \item[$\mathbf{q_2}$:]
    This gives the halo mass dependency of the screening scale.
    \item[$\mathbf{q_3}$:]
    This gives the environment dependency of the screening scale. 
    \item[$\mathbf{q_4}$:]
    This calibrates any existing Yukawa suppression scale. 
\end{itemize}
The time dependence of $\mathcal{F}_{\rm Erf}$ is fixed and so for a specified cosmology and set of EFTofDE parameters, we only need to adjust the constants $\{q_1, q_2, q_3,q_4\}$. To provide some insight, we note the following limits
\begin{align}
    \lim_{q_1 \rightarrow \infty} 1+ \mathcal{F}_{\rm Erf} & = \mu  \rightarrow  {\rm \bf Unscreened \, limit} \, ,  \\ 
    \lim_{q_1 \rightarrow -\infty} 1 + \mathcal{F}_{\rm Erf} & = 1  \rightarrow {\rm \bf GR \, limit}  \, , \\ 
    \lim_{q_2, q_3 \rightarrow 0} 1 + \mathcal{F}_{\rm Erf} & \rightarrow   {\rm \bf Vainshtein \, type \, models} \, ,  \\
    \lim_{q_3 \rightarrow 0} 1 + \mathcal{F}_{\rm Erf} & \rightarrow  {\rm \bf k-mouflage \, type \, models} \, , \\ 
     q_3>0: \,  1 + \mathcal{F}_{\rm Erf} & \rightarrow  {\rm \bf chameleon \, type \, models} \, ,
\end{align}
where we refer to the main types of screening mechanisms typical of scalar-tensor theories (see \autoref{sec:pffF}). Note that all parameters lose their meaning as $\mu(k,a) \rightarrow 1$, which in the EFTofDE case is when the relevant parameters assume their GR values.

Given this, we can take $q_2$ and $q_3$ to be positive. Being exponents of the top-hat radius and environment parameter, they are also not expected to be very large, and as we will see in \autoref{sec:validation}, they turn out to be $\mathcal{O}(1)$. Further, since in the GR limit $\mu \rightarrow 1$, and so  $\mathcal{F}_{\rm Erf} \rightarrow 0$ irrespective of the value of $q_1$, we can also take $q_1$ to be positive. We also find $q_1$ to be an $\mathcal{O}(1)$ parameter.

Parameter $q_4$, which calibrates the Yukawa suppression scale, is generally only relevant for theories where the linear growth factor, or Poisson modification $\mu$, is scale-dependent. As we will show in \autoref{sec:valfr}, $q_4$ does not appear to be relevant for the scales associated with spherical collapse. We note $q_4$ can in principle take on negative values, pushing the Yukawa suppression to smaller scales. As $q_4\rightarrow \infty$ the Yukawa suppression scale also goes to infinity. We leave its relevance for more general theories for a future work.   

\bigbreak

We provide a {\tt Mathematica} notebook, \href{https://github.com/nebblu/ACTio-ReACTio/tree/master/notebooks}{{\tt Nonlinear.nb}}, with all the forms of $\mathcal{F}$ considered in this paper along with comparisons.
\newline
\newline 
Finally, the left half of \autoref{fig:pipeline} summarises the map from the parametrised action, together with additional parameters, to the Poisson equation modifications as described in this section, completing the map from action to reaction. 


\section{Approximations and Overview}\label{sec:approximations}

We have outlined a map that goes from the parameterised action of nature and structure formation $ \{ H \}_b$, $\{\alpha_M, \alpha_B, \alpha_K, \alpha_T \}_{\rm L}$, $\{ \xi_3^2, \xi_4^2, \xi_3^3, \xi_4^3, \xi_4^4, \xi_4^5 \}_{\rm QNL}$ \& $\{ p_1,p_2,p_3,p_4,p_5,p_6,p_7,p_8 \}_{\rm NL}$ or $\{q_1,q_2,q_3,q_4 \}_{\rm NL}$  to the nonlinear effects on the power spectrum $\mathcal{R}(k,a)$, where `b' stands for background, `L' for linear, `QNL' for quasi-nonlinear and `NL' for nonlinear. A schematic of this map is given in \autoref{fig:pipeline}. An important point worth stressing is that our nonlinear parametrisations are completely general, and not specific to the Horndeski class of theories. They do however rely on $\mu(k,a)$, which one can always choose to parametrise in a model independent way. 

Considering the Horndeski class for concreteness, the EFTofDE and nonlinear parametrisations constitute a very large set of arbitrary functions of time and constants. Despite it being significantly less than the infinite number of theories contained within the Horndeski class, it is still arguably too many for statistical data analyses, both on computational and scientific grounds. Thankfully, as we will shortly motivate, these sets can be yet reduced significantly.

To reduce or optimise the parameter space, we consider the following: 
\begin{enumerate}
    \item[({\bf 1})]
    We assume the quasi-static approximation (QS) for all perturbative calculations \citep[see][for example]{Sawicki:2015zya,Pace:2020qpj}.
     \item[({\bf 2})]
    We assume $\gamma_2 = \gamma_3 = 0$. 
   \item[({\bf 3})]
    Observational and theoretical constraints.
    \item[({\bf 4})]
    Time parameterisations of EFTofDE functions, $\alpha_i(a)$. 
    \item[({\bf 5})]
    The parameterised nPPF (see \autoref{eq:nPPFF}) or phenomenological (see \autoref{eq:ErfF}) form of $\mathcal{F}$ is flexible enough to capture general modifications to gravity. 
\end{enumerate}
In this section we will motivate approximations {\bf (1)} - {\bf (4)} with direct reference to the accompanying code \href{https://github.com/nebblu/ACTio-ReACTio}{\tt ACTio et ReACTio}. Assumption {\bf (5)} will be addressed separately in \autoref{sec:validation}.


\subsection{Quasi-static approximation}

We begin by noting that the QS in linear theory can be easily avoided by using a Boltzmann code such as {\tt EFTCAMB} \citep{Hu:2013twa,Raveri:2014cka} to calculate the linear input spectrum or transfer function\footnote{The QS can be partly circumvented in the nonlinear regime, \autoref{eq:nPPFF} and \autoref{eq:ErfF}, by also using the prediction of $\mu(k,a)$ taken from say {\tt EFTCAMB}. }. This option is available in our code, but the default setting assumes a $\Lambda$CDM linear spectrum or transfer function at $z=0$ and rescales it using the internally calculated growth functions of the desired theory. This is done using the linear form of \autoref{eq:poisson1} (see \autoref{eq:mueft}) which assumes the quasi-static approximation. Being able to use a $\Lambda$CDM linear spectrum enhances the computational efficiency of our code as it avoids a call to {\tt EFTCAMB}. {\tt EFTCAMB} is significantly slower than {\tt CAMB} \citep{Lewis:1999bs}, which already takes $\mathcal{O}(1)$ seconds to produce a linear spectrum. In this case one can also use a linear spectrum emulator like {\tt CosmoPower} \citep{SpurioMancini:2021ppk} or {\tt bacco} \citep{Arico:2021izc}, which takes $\mathcal{O}(0.1)$ seconds to produce the linear spectrum. Note that one can also employ {\tt CosmoPower} to construct an emulator for the linear power spectrum in the EFTofDE based on {\tt EFTCAMB} output, overcoming the QS and computational inefficiency issues.

\begin{figure*}
    \centering
    \includegraphics[width=0.45\textwidth]{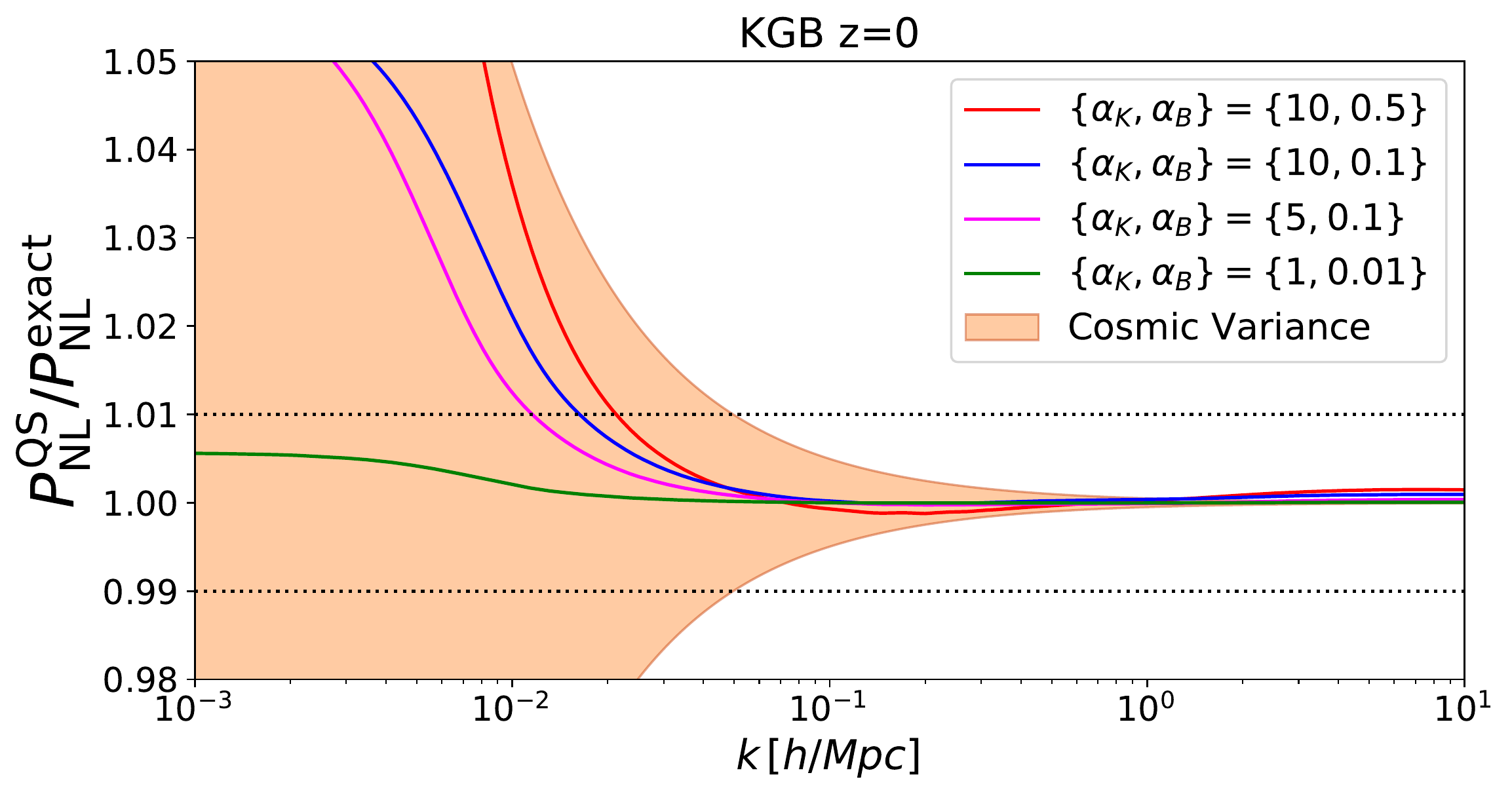} 
    \includegraphics[width=0.45\textwidth]{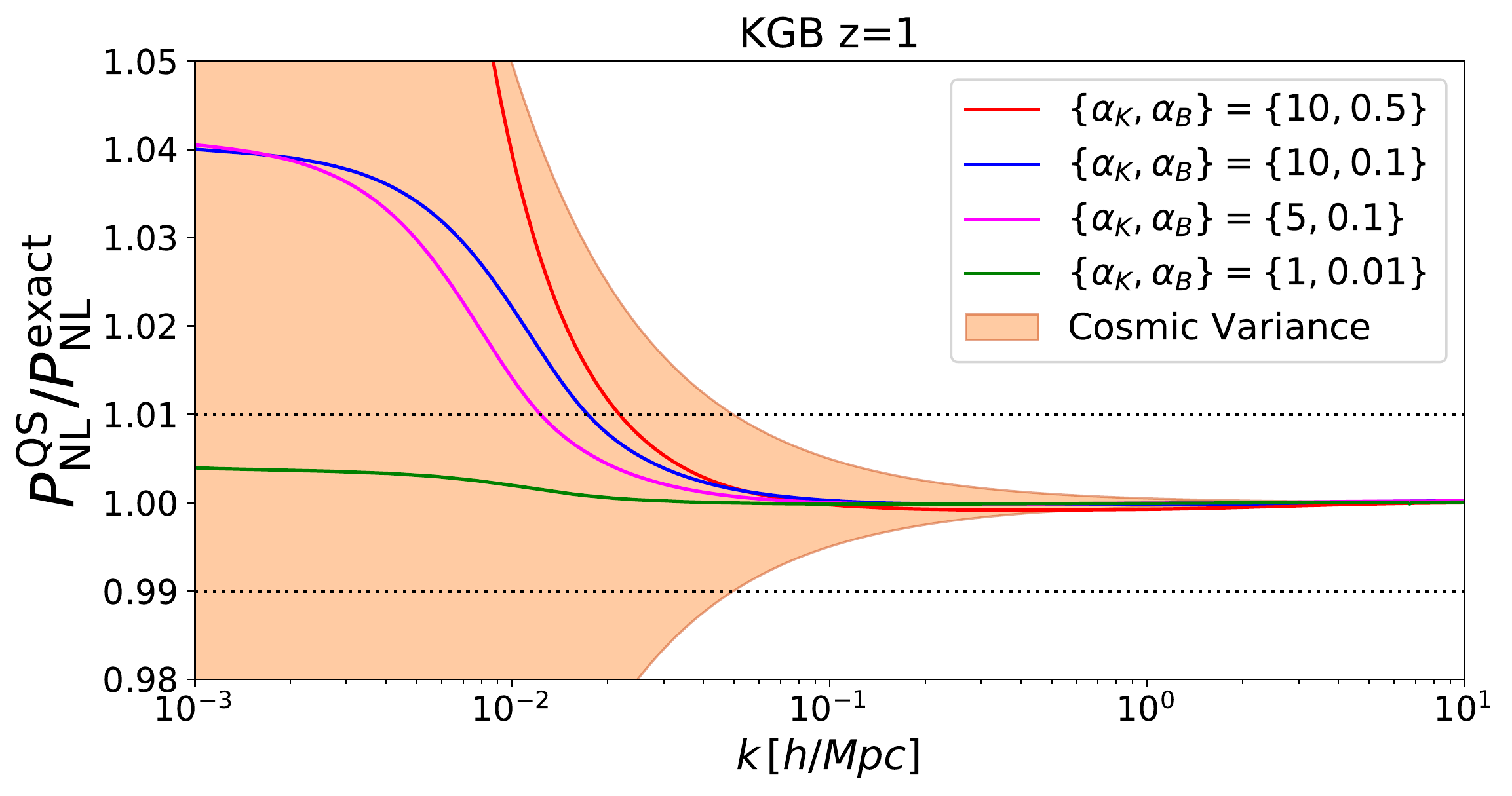}
    \caption{Ratio of the quasi-static approximated (QS) nonlinear spectrum to the exact calculation. We show the results for four EFTofDE models with $\{\alpha_K,\alpha_B\}$ non-zero and all other $\alpha$ parameters set to 0 and a $\Lambda$CDM background expansion, at $z=0$ ({\bf left}) and $z=1$ ({\bf right}). 
    The exact calculation uses  \autoref{eq:reaction} with an {\tt EFTCAMB} linear spectrum while the QS uses \autoref{eq:reaction} with a rescaled $\Lambda$CDM linear spectrum using the modified growth equations, making use of \autoref{eq:mueft}. The orange band represents the error coming from  cosmic variance assuming an effective survey volume of $V_{\rm eff} = 20 ~{\rm Gpc}^3 / h^{3}$. We assume $\mathcal{F} = \mu-1$ and $\mathcal{E}=1$ in all nonlinear computations. The dotted lines mark 1\% deviations which is an optimistic estimate on the modelling errors of the halo model reaction framework.}
    \label{fig:quasistat}
\end{figure*}

Given the utility in using the QS, we want to get an idea of its validity. In \autoref{fig:quasistat} we show the effects of the QS at $z=0$ and $z=1$ for models with non-zero $\alpha_K$ and $\alpha_B$  \citep[KGB][]{Deffayet:2010qz}, on the nonlinear spectrum as given by \autoref{eq:nonlinpk}. We use the halofit \citep{Takahashi:2012em} formula for $P_{\rm NL}^{\rm pseudo}$ and assume a $\Lambda$CDM background expansion, $H(a) = H_{\rm \Lambda CDM}(a)$ as well as no screening effects, i.e., $\mathcal{F} = \mu -1$ and $\gamma_2 = \gamma_3 = 0$.  

We find that the QS is valid for these mild to moderate parameter choices on scales of $k\geq 0.1 h/{\rm Mpc}$. Upcoming surveys will probe scales larger than this which may be an issue. Taking into account cosmic variance assuming a galaxy survey volume similar to the effective volume of forthcoming surveys, $V_{\rm eff} = 20 ~{\rm Gpc}^3 / h^{3}$ \citep{Laureijs:2011gra,Aghamousa:2016zmz,Euclid:2019clj}, the QS is still a sub-dominant source of error for even extreme choices of $\alpha_B$ and $\alpha_K$ (see \autoref{sec:constraints}). Note that time derivatives of the fields drop out from the calculation of $\mu$ for $k\rightarrow \infty$ in Horndeski theories \citep{Lombriser:2015cla,Pace:2020qpj}. We note at small scales, modelling inaccuracies and shot noise errors will arguably dominate any inaccuracies incurred from using the QS. 

We do however warn that the QS begins to break significantly for beyond Horndeski theories \citep{Lombriser:2015cla}. For large modifications to GR within Horndeski, we advise comparing the resulting nonlinear spectrum with and without the QS against the predicted errors on the specific data that is being analysed. Further, we have implemented the following necessary condition for the QS to hold in our code \citep{Peirone:2017ywi}
\begin{equation}
    \frac{k}{a H(a)} > c_s^2(a) \, , 
\end{equation}
where $c_s^2$ is given by \autoref{eq:cs2}, with its violation producing a warning prompt. 


\subsection{$\gamma_2 = \gamma_3 = 0$ approximation}

We begin by noting that setting $\gamma_2=\gamma_3=0$ implies we have $R_{\rm SPT}\approx 1$ in \autoref{eq:T1} as the 1-halo terms are subdominant. This forces the argument of the logarithm in \autoref{kstar} to be very close to unity, giving a very large $k_{\rm star}$. Effectively, this is the same as setting $\mathcal{E} = 1$ in \autoref{eq:1hcb}. This is the choice we take when adopting this approximation. We should remark that simply setting $\gamma_2=\gamma_3=0$ leaves one slightly sensitive to the $k_{\rm star}$ correction through the 1-halo terms and consequently on the particular choice of halo mass function.

Using the exact forms of $\gamma_2$ \& $\gamma_3$ as described in \autoref{sec:eftofdeqsa} is a big challenge. This is primarily for computational reasons as it involves numerical time derivatives. Smoothness of such derivatives is difficult to ensure and can affect results. In particular, the exponential dependence of $\mathcal{R}$ on $k_\star$ (see \autoref{eq:reaction}) makes it very sensitive to inaccuracies in the 1-loop calculation. Further, the full map to $\gamma_2$ \& $\gamma_3$ from the EFTofDE would increase computational time significantly, degrading our code's ability to perform statistical analyses on data. 

To test the impact of setting $\gamma_2 = \gamma_3 = 0$ we compare \autoref{eq:reaction} with and without these terms switched on for two different theories of gravity, DGP and the Hu-Sawicki $f(R)$ model \citep{Hu:2007nk}. The former is an instance of derivative or Vainshtein screening and the latter of potential or chameleon screening, covering two main types of screening mechanism. 

\begin{figure*}
    \centering
    \includegraphics[width=0.45\textwidth]{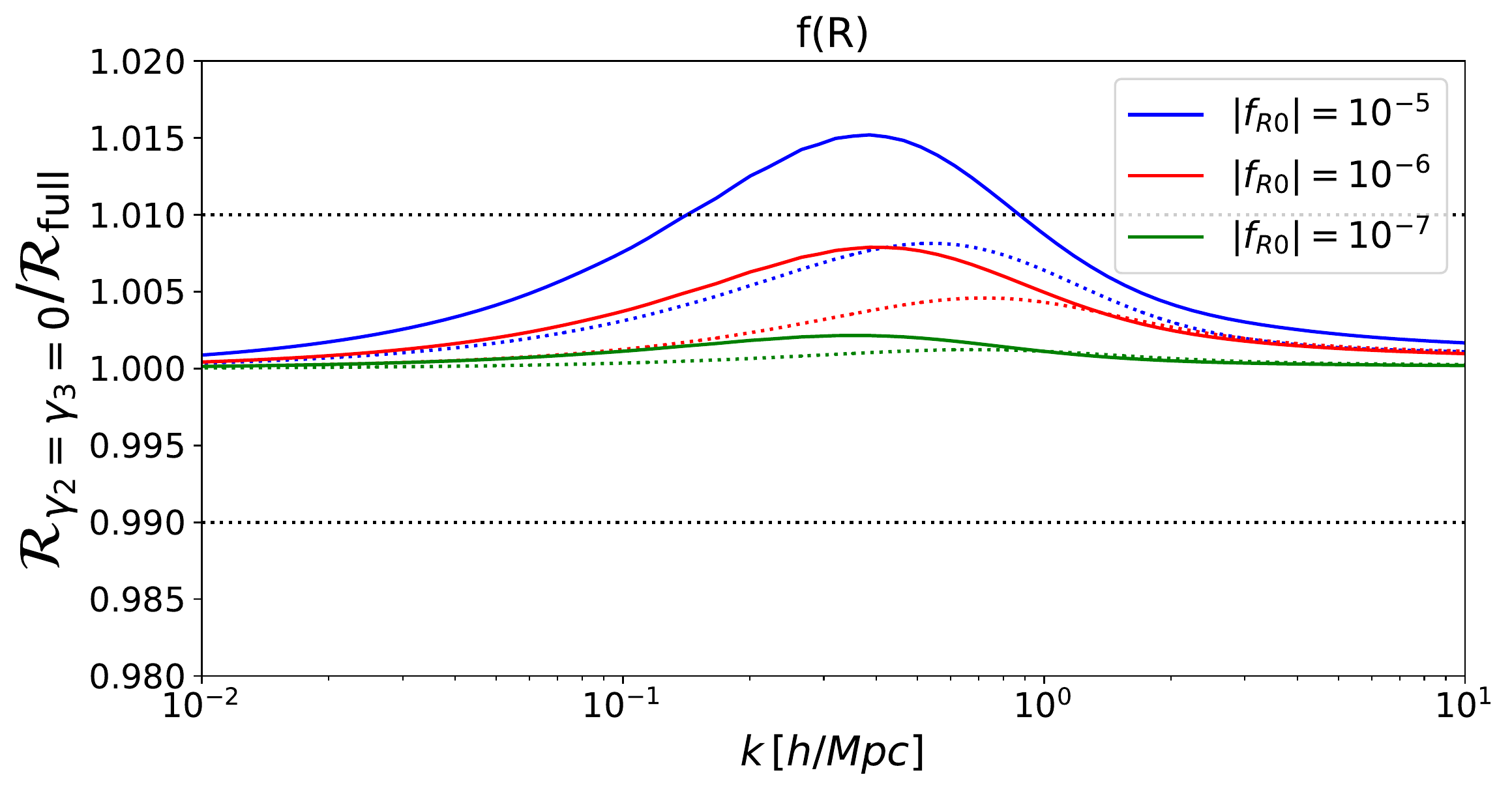}
    \includegraphics[width=0.45\textwidth]{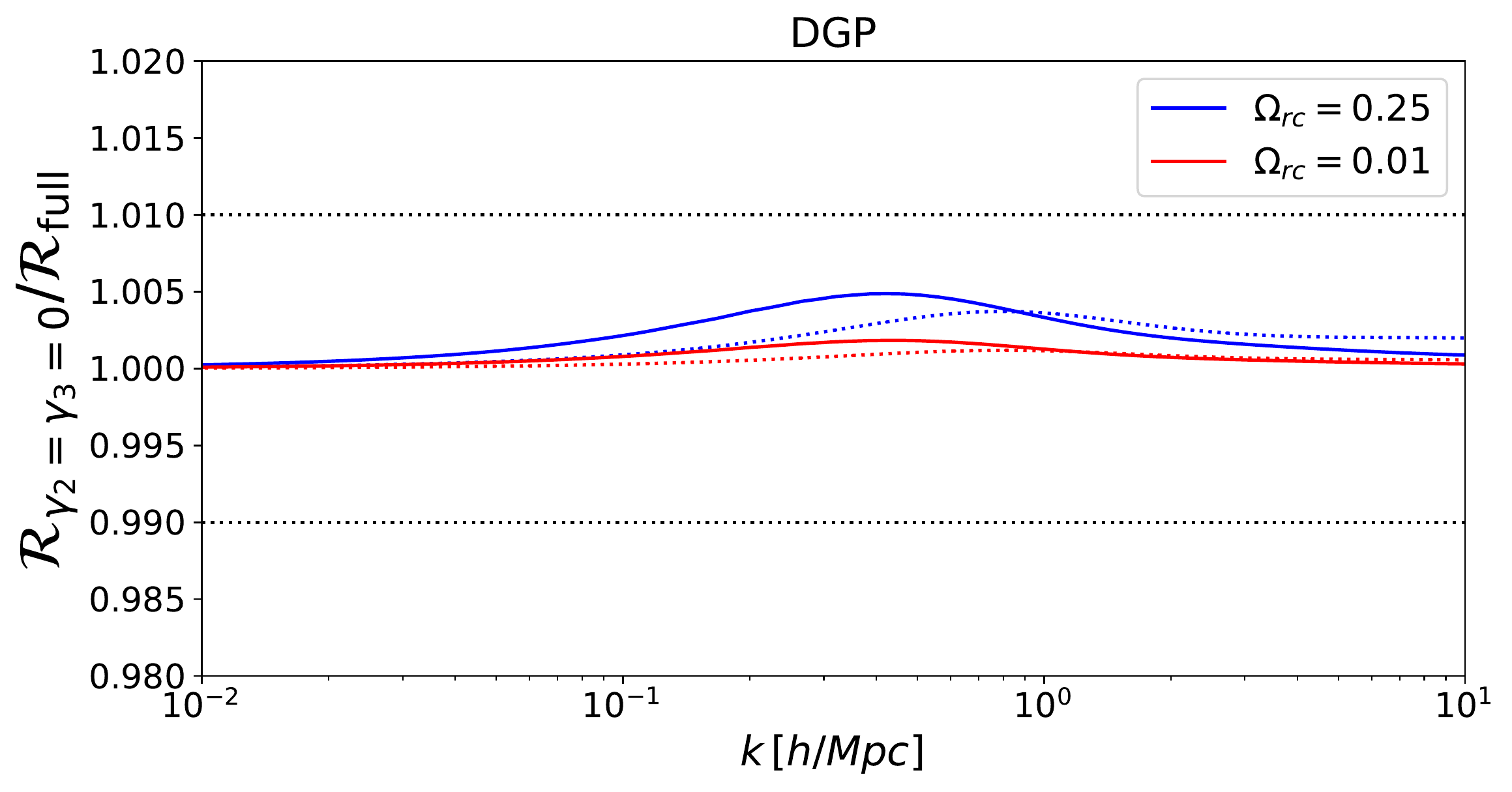}
    \caption{Ratio of the approximated reaction to the full calculation. The approximation assumes $\mathcal{E}=1$ in \autoref{eq:1hcb} which is approximately equivalent to no higher order perturbative, screening terms i.e, $\gamma_2=\gamma_3=0$. We show Hu-Sawicki $f(R)$ gravity on the {\bf left} and the normal branch of DGP on the {\bf right} for varying modifications to GR. For $f(R)$ we show predictions for when the model parameter takes the value of $|f_{R0}| = 10^{-5}$ (moderate modification, {\bf blue}), $|f_{R0}| = 10^{-6}$ (low modification, {\bf red}) and $|f_{R0}| = 10^{-7}$ (very low modification, {\bf green}). For DGP we show predictions for when the model parameter is $\Omega_{\rm rc} = 0.25$ (moderate modification, {\bf blue}) and  $\Omega_{\rm rc} = 0.01$ (low modification, {\bf red}). We also show the comparison over two redshifts, $z=0$ ({\bf solid lines}) and $z=1$ ({\bf dotted lines}).}
    \label{fig:reactiontest}
\end{figure*}

This comparison is shown in \autoref{fig:reactiontest}. We find that in the case of DGP, the correction coming from the 1-loop computation is negligible for small and moderate modifications to GR at all scales. On the other hand, the corrections to the $f(R)$ theory can be up to $1.5\%$ at $z=0$ for moderate modifications to GR. This may be acceptable if these inaccuracies can be partially absorbed into the nonlinear degrees of freedom. We explore this in \autoref{sec:validation}.


\subsection{Observational and theoretical constraints}\label{sec:constraints}

\begin{table*}
\centering
\caption{Theoretical and observational constraints on $\alpha$-parameters from references in the right column. Theoretical constraints are coming from low-energy (EFT) and high-energy (positivity bounds) physics. Note that $\alpha_K$ is not constrained by data, hence the subluminality condition does not impose any constraining power on the scalar mode perturbations. Also note the contradiction in the condition for GW propagation: subluminal versus superluminal speed. The positivity bounds do not hold in general, they are derived for a quadratic subclass of Horndeski theories with $G_3 = G_5 = G_{5,X} = 0$ in \autoref{eq:horndeski}. Data driven constraints strongly depend on the imposed theoretical priors and time-dependent parametrisation of $\alpha$-functions. Here we mention only two prior-independent observational constraints.}
\begin{tabular}{| c | c | c | c | c | }
  & & {\bf scalar} & {\bf tensor} & \\ \hline 
  & no ghost & $\alpha_K + \frac{3}{2} \alpha_B^2 >0$ & $M^2>0$ & \multirow{2}{*}{\cite{Bellini:2014fua}} \\ 
  Low & gradient stability & $c_s^2 \geq 0$ & $\alpha_T \geq -1$&\\
  Energy & (sub)luminality & large $\alpha_K$  & $\alpha_T \leq 0$& \cite{deRham:2018red} \\
  & no GW-induced instability & $| \alpha_M + \alpha_B| \lesssim 10^{-2}$  & & \cite{Creminelli:2020}\\ \hline

  High & scalar-scalar scattering & \multicolumn{2}{c}{$\alpha_B \leq \frac{2 \alpha_T}{1+\alpha_T}$} & \cite{Melville:2019wyy}\\
  Energy & scalar-matter scattering& \multicolumn{2}{c}{$\alpha_T \geq 0$} & \cite{deRham:2021fpu}\\ \hline \hline
  Data & GW propagation speed & \multicolumn{2}{c}{$| \alpha_T | \leq 10^{-15}$} & \cite{Monitor:2017mdv}\\
  & CMB and LSS & \multicolumn{2}{c}{$| \alpha_M |, | \alpha_B | \leq \mathcal{O}(0.1)$} & \cite{SpurioMancini:2019rxy}

\end{tabular}
\label{tab:theorpriors}
\end{table*}
Firstly, we want to eliminate a range of $\alpha$-parameter values that leads to two pathological instabilities: ghost (i.e., negative kinetic energy) and gradient (i.e., imaginary speed of sound). These constraints for the Horndeski theories were first derived in \citet{DeFelice:2011bh}.  In terms of the $\alpha$-functions, \citet{Bellini:2014fua} found that the stability of the background requires
\begin{equation}
    \alpha > 0 \, , \,\,\,\,\, c_s^2 \geq 0 \, ,
\end{equation}
from \autoref{eq:cs2} and \autoref{eq:alpha} for scalar modes, and 
\begin{equation}
    M^2 > 0 \, , \,\,\,\,\, c_T^2 = 1 + \alpha_T \geq 0 \, , 
\end{equation}
for tensor modes of perturbations. An additional theoretical constraint is the stability of scalar modes in the presence of gravitational waves of large amplitude, for instance, sourced by massive binary systems \citep{Creminelli:2020}. Mapped to the parameterisation used in this work this requires the following bound \citep{Noller:2020afd}:
\begin{equation}
    |\alpha_M + \alpha_B| \lesssim 10^{-2} \, .
\end{equation}
Previously, it was argued that the constraining power of  upcoming cosmological surveys will allow us to pin down the $\alpha$-parameters  at the $\mathcal{O} (0.1)$-level \citep[e.g., ][]{Frusciante:2018jzw}. For the condition above this implies that $\alpha_M \approx -\alpha_B$. However, in such forecasts nonlinear scales were ignored with a typical highest mode around $k_\mathrm{max} \approx 0.15$ $h$ Mpc$^{-1}$. We speculate that this constraint may be improved upon by inclusion of the nonlinear scales. Therefore, in our code we treat $\alpha_B$ and $\alpha_M$ independently.

Secondly, one may consider that the new physics should not modify the speed of gravitational wave propagation \citep{Lombriser:2015sxa,Monitor:2017mdv,Lombriser:2016yzn,Creminelli:2017sry,Ezquiaga:2017ekz,Baker:2017hug,Sakstein:2017xjx,Battye:2018ssx,deRham:2018red,Creminelli:2018xsv}, and so $\alpha_T = \bar{M}_2^2 = 0$. This luminality condition has been argued to not be as clear cut a constraint through EFT considerations \citep{deRham:2018red,LISACosmologyWorkingGroup:2022wjo} as well as through the positivity bounds from high energy physics \citep{deRham:2021fpu}, so in our code we keep the $\alpha_T$ dependence in $\mu$. Subluminality, stated in the former references, follows from the existence of a Wilsonian UV completion \citep{Adams:2006sv} and dependence on the theory's `cutoff' scale. From \autoref{eq:cs2} it can be seen that subliminality of scalar modes is guaranteed for large values of $\alpha_K$, while for tensor modes subluminality requires $\alpha_T < 0$.  Superluminality, stated in \citet{deRham:2021fpu}, is a consequence of the positivity bounds for scattering between scalar and matter fields. Such positivity bounds require a unitary, causal, local UV completion of our low-energy EFT theory. However, superluminality does not necessary result in casual paradoxes \citep{Babichev:2007dw, Burrage:2011cr}. In general, the notion of causality in terms of the low-energy EFT is a rather subtle topic \citep[for instance, see][]{deRham:2020zyh,Reall:2021voz}.

Thirdly, in the QS $\alpha_K$ does not enter the equations of motion \citep{Bellini:2014fua}. Therefore, it is completely unconstrained in our approach, or for any model with $c^2_s \approx 1$. However, in the exact computation $\alpha_K$ affects only the largest scales (see \autoref{fig:quasistat}), which are dominated by cosmic variance. This can be a motivation to not consider $\alpha_K$ in data analyses, leaving only $\alpha_M$ and $\alpha_B$ in a `bare-bones' case. We do not impose any of these reductions in our code and leave it to the user to specify well motivated priors on the full set of EFTofDE parameters in their analyses.   

Lastly, we note that there are a host of data driven constraints that one can put on the EFTofDE parameters \citep{Huang:2015srv,Bellini:2016,Noller:2018wyv,Noller:2019,SpurioMancini:2019rxy,Melville:2019wyy,Noller:2020afd,deRham:2021fpu}. Such constraints strongly depend on the imposed theoretical priors and time-dependent parameterisation of the $\alpha$-functions (see \autoref{sec:timedepalpha}). However, they all agree that the uncertainties and values of the $\alpha$-parameters are of order $\mathcal{O}(0.1)$. The future CMB and LSS surveys promise to improve the constraints up to at least one order of magnitude $\sigma(|\alpha_i|) \sim \mathcal{O}(0.01)$ \citep[see, for example,][]{CMB-S4:2016ple}. One may also assume a $\Lambda$CDM background, well motivated by CMB data \citep[e.g.,][]{Planck:2018vyg}, and so set $H(a) = H_{\rm \Lambda CDM}$(a)\footnote{Our code defaults to this assumption, but there is the option to parameterise the background too.}. We summarize the constraints discussed above in \autoref{tab:theorpriors}.


\subsection{Parameterising time dependence}\label{sec:timedepalpha}

Here, we look at how one can parameterise the time dependence of the EFTofDE functions. To first order this can be approximated by a Taylor expansion, $\theta_i(a) \approx \theta_{i,0} + \theta_{i,p} (1-a)$,  leaving at least 6 free constants characterising deviations from $\Lambda$CDM. In typical data analyses, only a 1-parameter time dependence is considered. For example, in \cite{Noller:2018wyv} the authors consider the following three parameterisations for the $\alpha_i$, $i \in \{M,B,K,T\} $
\begin{align}
   \mathbf{(1):} \,  \alpha_i(a) & = c_i \Omega_{\rm \Lambda}(a) \, , \\ 
   \mathbf{(2):} \,  \alpha_i(a) & = c_i a \, , \\ 
  \mathbf{(3):} \,   \alpha_i(a) & = c_i a^{n_i} \, , 
\end{align}
where $c_i$ and $n_i$ are free constants and $\Omega_{\rm \Lambda}(a)$ is the $\Lambda$CDM cosmological constant energy density fraction as a function of time. For a comprehensive list of various other time parameterisations see Appendix B of \cite{Frusciante:2019xia}. These all draw on the motivation that modifications should only become relevant at late times. In our code, the default is set to {\bf (2)} for all $\alpha_i$. We note that such parametrisations may exclude well-known theories as shown in \cite{Kennedy:2018gtx}, which motivated the $s$-basis introduced in \autoref{sec:eftofde}.

We can also adopt similar parametrisations for the background $H(a)$, but a more general choice would be for example the Chevalier-Polarski-Linder (CPL) parametrisation \citep{Chevallier:2000qy,Linder:2002et}, which parametrises the dark energy equation of state $w(a)$ in terms of two free constants, $\{w_0,w_a\}$ as 
\begin{equation}
    w(a) = w_0 + w_a(1-a) \, , 
\end{equation}
which gives the following form for $H(a)$
\begin{equation}
H^2(a) = H^2_0\left(\Omega_{{\rm m},0} a^{-3} + \Omega_{\Lambda} e^{-3\int(1+w(a)){\rm d} \log a} \right)\, .
\label{eq:cpl_ha}
\end{equation}
\begin{table*}
\centering
\caption{A maximal, reduced and minimal set of parameters needed for a comprehensive nonlinear power spectrum analysis of the unrestricted theory space of \autoref{eq:horndeski}, together with a proposed minimal set for more general theories. The Horndeski minimal set assumes $\alpha_T = \alpha_K = 0$ and \autoref{eq:Erfexp}, while the maximal and the reduced assume full freedom of \autoref{eq:nPPFF}. Both reduced and minimal assume $\gamma_2=\gamma_3=0$. The totals show the number of free functions of time plus any free constants. We note $q_4$ and $p_8$ are found in \autoref{sec:validation} to be likely irrelevant for the 1-halo computation, and so we do not consider them in the reduced or minimal cases. We also note very tight constraints on $\alpha_B-\alpha_M$ (see \autoref{sec:constraints}) relevant to the minimal case. } 
\begin{tabular}{| c | c | c | c | c | }
  & {\bf Maximal} & {\bf Reduced} & {\bf Minimal (Horndeski)} & {\bf Minimal (general)}\\ \hline 
{\bf Background} &  $H(a)$ &  $H(a)$ & $H(a)$ & $w_0,w_a$ \\ 
{\bf Linear} & $  \alpha_M (a), \alpha_B(a), \alpha_K(a), \alpha_T (a) $ & $  \alpha_M (a), \alpha_B(a), \alpha_K(a), \alpha_T (a) $ &  $ \alpha_M (a),  \alpha_B(a) $ & $\gamma$  \\ 
{\bf Quasi-nonlinear} & $\xi_3^2(a), \xi_4^2(a), \xi_3^3(a),\xi_4^3(a), \xi_4^4(a), \xi_4^5(a)$ & -  & - & -  \\ 
{\bf Nonlinear} & $ p_{1-7}(a)+p_8  $  & $p_{1-7}(a)  $ & $q_1,q_2,q_3$  & $q_1,q_2,q_3$  \\ \hline
{\bf {Total}} & 18+1 & 12  & 3 + 3 constants & 6 constants
\end{tabular}
\label{tab:parameterisations}
\end{table*}


\subsection{Parametrisation of $\mathcal{F}$}
The nPPF form for $\mathcal{F}$ given in \autoref{eq:nPPFF} captures dependencies of the nonlinear modification to the Poisson equation on the relevant variables, namely $\{y_{\rm h}, a, M_{vir}, y_{\rm env}\}$. Being motivated by the form of $\mathcal{F}$ in DGP (\autoref{eq:nlexactdgp}), it can recover the DGP form given appropriate choices for $p_i$ albeit with a non-trivial dependency of $p_2$ on $a$ (see \autoref{eq:dgppi}). \autoref{eq:nPPFF} becomes approximate when moving beyond DGP. On the other hand, the Erf form, \autoref{eq:ErfF}, is completely phenomenological and is an approximation even in DGP. 

Note that the nPPF is also more directly relatable to specific actions and gravity models, in which case its degrees of freedom can be significantly restricted. It is thus far more suitable when particular models are being targeted for analysis. The Erf model on the other hand is completely general and has no direct relation to specific actions of gravity. It is thus more suitable when no specific model is being targeted and we want to place constraints on general models of gravity. In \autoref{sec:validation} we test these two approximations in both DGP and $f(R)$ gravity.

\subsection{Overview} 

With all these approximations and constraints, the arguable minimal parameter space characterising deviations to $\Lambda$CDM is 3 free functions of time and 4 constants. Without approximations or constraints, the maximal is 18 free functions of time and a constant. Of course we can also find intermediate reduced sets, such as using the nPPF but with $\gamma_2=\gamma_3=0$. Given we need to parameterise these functions of time, the maximal set is currently an unfeasible parameter space to probe comprehensively, both in terms of data processing as well as parameter degeneracies which limits the amount of useful physical information one can extract from the data.

Finally, we have focused on the Horndeski class of models, but one can extend this to larger generality by considering for example the growth index $\gamma$ parametrisation for $\mu(a)$ \citep{Peebles1980,Linder:2007hg} \citep[explicitly, see Eq.~47 of][]{Kennedy:2018gtx} and \autoref{eq:cpl_ha} for $H(a)$. Combined with the Erf model, this would constitute a minimal set of 6 free constants for general modifications to $\Lambda$CDM. This minimal model-independent parametrisation has also been implemented into the code.

We summarise  these parameterisations in \autoref{tab:parameterisations}.


\section{Testing the nonlinear parameterisations} \label{sec:validation}

In this section we compare the predictions for the halo model reaction $\mathcal{R}$, using the various nonlinear parameterisations of modifications to the Poisson equation outlined in \autoref{sec:nonlinpar}, to exact solutions as well as state-of-the-art emulators within an evolving dark energy scenario (wCDM), DGP and Hu-Sawicki $f(R)$ gravity. Note that the exact solutions for the reaction have in turn been themselves compared to full $N$-body simulations in other works \citep[see][for example]{Cataneo:2018cic}, exhibiting $\mathcal{O}(1)\%$ agreement. These models cover a fair range of theoretical and phenomenological features typical of modified gravity and dark energy models, making them good representatives and test cases. 

We look to test predictions for $\mathcal{R}$ using \autoref{eq:nPPFF} (nPPF) and \autoref{eq:ErfF} (Erf) with $\gamma_2=\gamma_3=0$ against the full calculation which computes $\mathcal{R}$ using exact forms for $\gamma_2$, $\gamma_3$ and $\mathcal{F}$ \citep[see Appendices of][for all relevant expressions]{Bose:2020wch}. We further employ the EuclidEmulator2 emulator \citep{Euclid:2020rfv} and the fofr emulator \citep{Winther:2019mus} for the wCDM and $f(R)$ cases respectively. These emulators have been trained on high quality $N$-body simulations and are 1-2\% accurate within the scales we examine, providing a good benchmark for our predictions. One should keep in mind that the halo model reaction approach's accuracy is limited by the pseudo power spectrum employed. For example, if we use {\tt HMCode2020} \citep{Mead:2020vgs} for the pseudo, which is claimed to be $2.5\%$ accurate down to $k\leq 10~h/{\rm Mpc}$, we then expect any power spectrum comparisons to then be $\sim4.5\%$ consistent with $N$-body at $k\leq 3~h/{\rm Mpc}$, which assumes the result of \cite{Cataneo:2016iav}, i.e., that the exact solution for $\mathcal{R}$ is $\sim 2\%$ accurate at these scales. In the wCDM and DGP cases, both $\Lambda$CDM and $P_{\rm NL}^{\rm pseudo}$ are computed using the halofit fitting function \citep{Takahashi:2012em}, but the $f(R)$ case uses {\tt HMCode2020}.

The computation of $\mathcal{R}$ requires us to solve the evolution equations for the spherical top-hat radius parametrised by $y_{\rm h}$ (\autoref{eq:ydef}). This necessitates the specification of $\mathcal{F}$ at all redshifts up to the target redshift. We then should test approximations for $\mathcal{F}$ even at high redshifts, which is done in \autoref{sec:nlfcomparisons}, where we compare $\mathcal{F}$ at $z=0,1,4$. For comparisons of the halo model reaction, we only consider $z=0,1$ which are more observationally relevant.

 We fit $\{q_1,q_2,q_3,q_4\}$ for the Erf model, $\mathcal{F}_{\rm Erf}$. In the nPPF case, we do not fit all the 8 free parameters of $\mathcal{F}_{\rm nPPF}$, and only consider $p_1$ and $p_8$, treating both as constants. In principle, and indeed for unspecified theories of gravity, all 8 parameters will be fit to the data. For the comparisons made here, $p_{2-7}$ are fixed to the theoretically predicted values quoted in \autoref{app:nonlin}. Fitting such a high dimensional parameter space is beyond the scope of this paper. 

In what follows we fit the free parameters by performing a least square fit to the exact $\mathcal{R}$ prediction. We choose to fit our parametrised models to the exact predictions for $\mathcal{R}$, rather than the emulator predictions for $P_{\rm NL}(k)$ for two reasons. First so as to test the ansatz for the phenomenological screening and the consistency of the predictions (see \autoref{app:nonlin}). Second, we do not want to assume anything about the pseudo spectrum in these fits. To fit we minimise the following merit function
\begin{equation}
    s^2 = \sum^{\rm max}_{j={\rm min}}\sum^{\rm max}_{i={\rm min}} \frac{[\mathcal{R}_{\rm exact}(k_i,z_j) - \mathcal{R}_{\rm approx}(k_i,z_j)]^2}{\sigma_{i,j}^2} \, , \label{eq:chi2} 
\end{equation}
where we assume error bars on $\mathcal{R}_{\rm exact}$ coming from cosmic variance \citep{Zhao:2013dza,Blanchard:2019oqi,Mancarella:2020jyu}  and a constant systematic error added in quadrature
\begin{equation}
    \sigma_{i,j}^2(k,z) = \frac{4\pi^2}{k_i^2 \Delta k_i V_{s,j}} + \sigma_{\rm sys}^2   \, , \label{eq:gaussianerr}
\end{equation}
where $V_{s,j} \in \{0.3,8\} ~ {\rm Gpc}^3/h^3$ is taken to be a stage IV survey-like volume for each bin $z_j \in \{0,1\} $ respectively  \citep{Laureijs:2011gra,Aghamousa:2016zmz,Mancarella:2020jyu,Euclid:2019clj}. We fit in the range $k_{\rm min} = 0.1 \leq k_i \leq 3 = k_{\rm max}$ which is the range over which the exact computation of $\mathcal{R}$ is $2\%$ accurate \citep{Cataneo:2018cic}, sampling logarithmically, with $\Delta k_i$ being the bin width. We take $\sigma_{\rm sys} = 0.02$ to reflect the systematic error in the parametrised reaction when compared to simulations by proxy of the exact solution. The best fit parameter values are shown in \autoref{tab:paramfits}.
\begin{table}
\centering
\caption{Best fit parameter values for the DGP and $f(R)$ models. The fit is performed to the exact solution for $\mathcal{R}$ in the range $0.1~h/{\rm Mpc}\leq k \leq 3~h/{\rm Mpc}$ and at $z=0,1$ as described in the main text. For the Erf model, we do not fit $q_2$, $q_3$ and $q_4$ for DGP and for $f(R)$ we find the quality of fit with and without $q_4$ is similar. For all fits we thus set $q_4=0$. The nPPF is exact for DGP and so we only consider $f(R)$, fixing all $p_{2-7}$ to the values given in \autoref{eq:frpi}. } 
\begin{tabular}{| c | c | c | c | c | c |  }
  & \multicolumn{2}{c}{{\bf nPPF}} & \multicolumn{3}{c}{{\bf Erf}} \\ \hline 
  $\Omega_{\rm rc} $ & $p_1$ & $p_8$ & $q_1$ & $q_2$ & $q_3$\\ \hline 
  0.25  & - & - & 0.76 & 0 & 0  \\ 
  0.01 & - & - & 0.71 & 0 & 0   \\  \hline 
  $|f_{\rm R0}|$ & $p_1$ & $p_8$ &  $q_1$ & $q_2$ & $q_3$  \\ \hline 
  $10^{-5}$  & 3 & -0.8 &  0.9 & 0.35 & 0.65   \\ 
  $10^{-6}$  & 8.5 & -0.5 &  1.65 & 0.7  & 2.45 \\ 
  $10^{-7}$  & 5.65 & -0.45 &  0.6 &  0.8 &  2.15    
\end{tabular}
\label{tab:paramfits}
\end{table}
%

\subsection{Evolving dark energy example: wCDM}
Here we perform a sanity check that the general minimal model outlined in \autoref{tab:parameterisations} produces consistent results for a wCDM cosmology, and is at least as accurate as the exact solution. To do this we compare a minimal model with CPL parameters $w_0=-1.2$ and $w_a=0.4$, and a growth index of $\gamma=0.55$ to the exact solution as well as predictions from EuclidEmulator2 using the same CPL parameters. We further set the nonlinear parameters of the Erf model ($q_i$) to unity, but check that they have no impact on the results as expected from \autoref{eq:ErfF} ($\mu \approx 1$ for $\gamma=0.55$).

\begin{figure}
    \centering
    \includegraphics[width=0.45\textwidth]{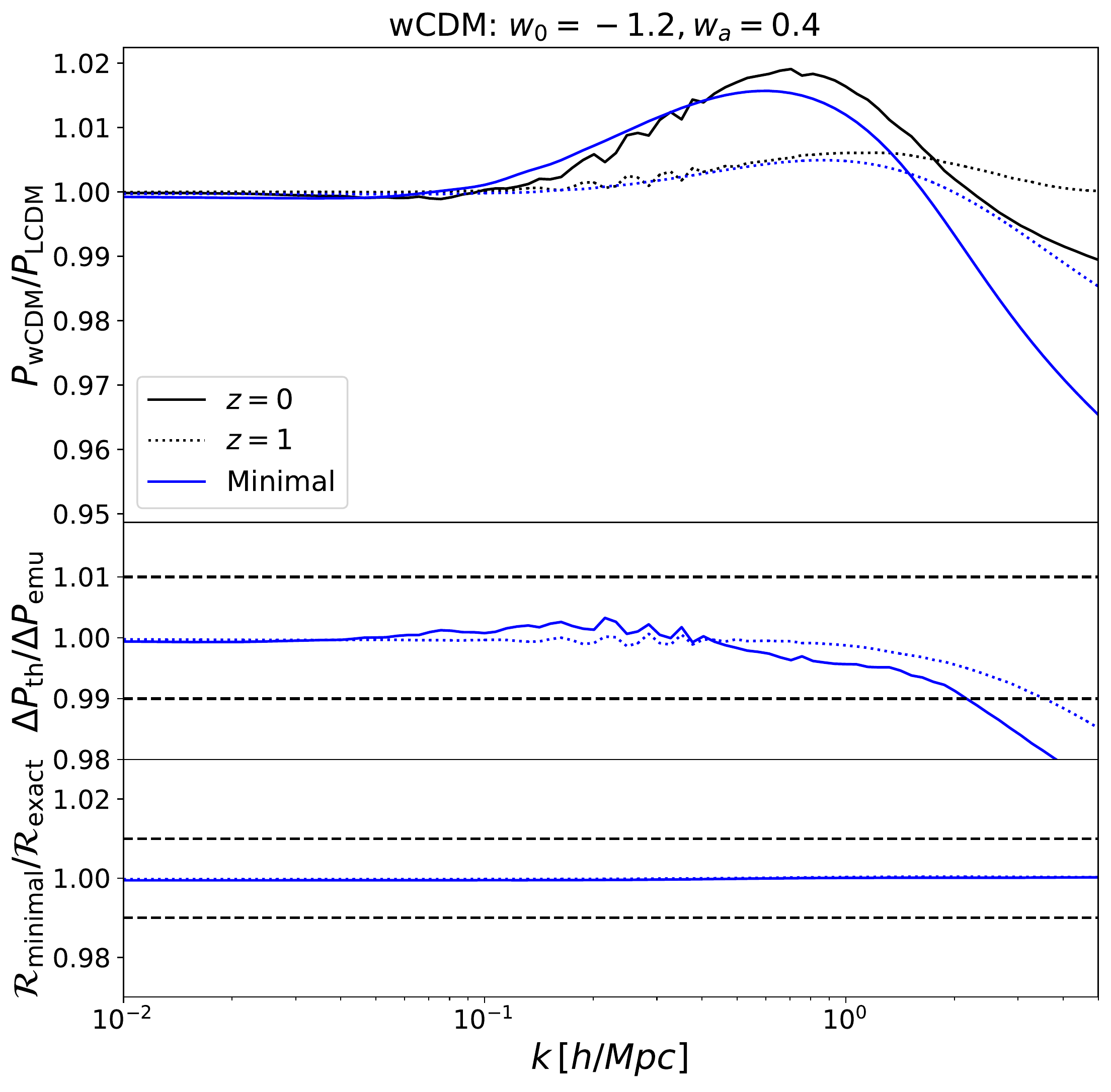}
    \caption{{\bf Top panel:} The ratio of the $w$CDM nonlinear power spectrum to the $\Lambda$CDM nonlinear power spectrum computed using the EuclidEmulator2 ({\bf black}) and halofit together with the halo model reaction (see \autoref{eq:nonlinpk}) for the minimal general model ({\bf blue}) as outlined in the right most column of \autoref{tab:parameterisations}. {\bf Middle panel:} The ratio between theoretical and emulator predictions for the ratio between wCDM to $\Lambda$CDM spectra, i.e., the ratio of blue to black top panels curves. {\bf Bottom panel:} The ratio of the exact halo model reaction to the minimal general model. We plot the ratio for two observationally relevant redshifts, $z=0$ ({\bf solid}) and $z=1$ ({\bf dotted}). We show these results for $w_0=-1.2$ and $w_a=0.4$. The minimal general model also has $\gamma=0.55$ and $q_1=q_2=q_3=q_4=1$. }
    \label{fig:wcdm_Rtest}
\end{figure}

We show our results in \autoref{fig:wcdm_Rtest}. We see that the minimal model is both completely consistent with the exact solution which has no nonlinear or linear modification to the Poisson equation, as well as $1\%$ consistent with the emulator down to $k\leq 2~h/{\rm Mpc}$ and $2\%$ down to $k\leq 3~h/{\rm Mpc}$. The minimal general model could feasibly outperform the exact solution given its degrees of freedom. In a future work we plan to check forecasted constraints and possible biases on cosmological parameters for the minimal general model, in full posterior estimation analyses employing $N$-body simulation measurements. 

\subsection{Vainshtein example: DGP}
For DGP the nPPF parameterisation reproduces the exact form of $\mathcal{F}$ (\autoref{eq:nlexactdgp}) for specific choices of the $p_i$ parameters (\autoref{eq:dgppi}). On the other hand, the Erf parametrisation (\autoref{eq:ErfF}) is approximate and we fit the associated parameters. We note that DGP has no Yukawa suppression at large scales and produces a constant enhancement of the $\Lambda$CDM linear growth factor. This enhancement is controlled by the DGP degree of freedom $\Omega_{\rm rc} \equiv 1/(4 H_0^2 r_c^2)$ where $r_c$ is the cross-over scale dictating where gravity goes from behaving 4-dimensionally to 5-dimensionally. We consider two levels of deviation to $\Lambda$CDM: a moderate modification given by $\Omega_{\rm rc}=0.25$ and a small modification given by $\Omega_{\rm rc}=0.01$.

We only fit $q_1$ as we do not have any mass, environment or Yukawa-suppression scale dependence, and so we set $q_2=q_3=q_4=0$ in this case. The best fit values of $q_1$ are given in \autoref{tab:paramfits}. Further, we employ the exact form of $\mu(a)$ in \autoref{eq:ErfF} \citep[see appendices of][for the explicit expression]{Bose:2020wch}.

\begin{figure*}
    \centering
    \includegraphics[width=0.45\textwidth]{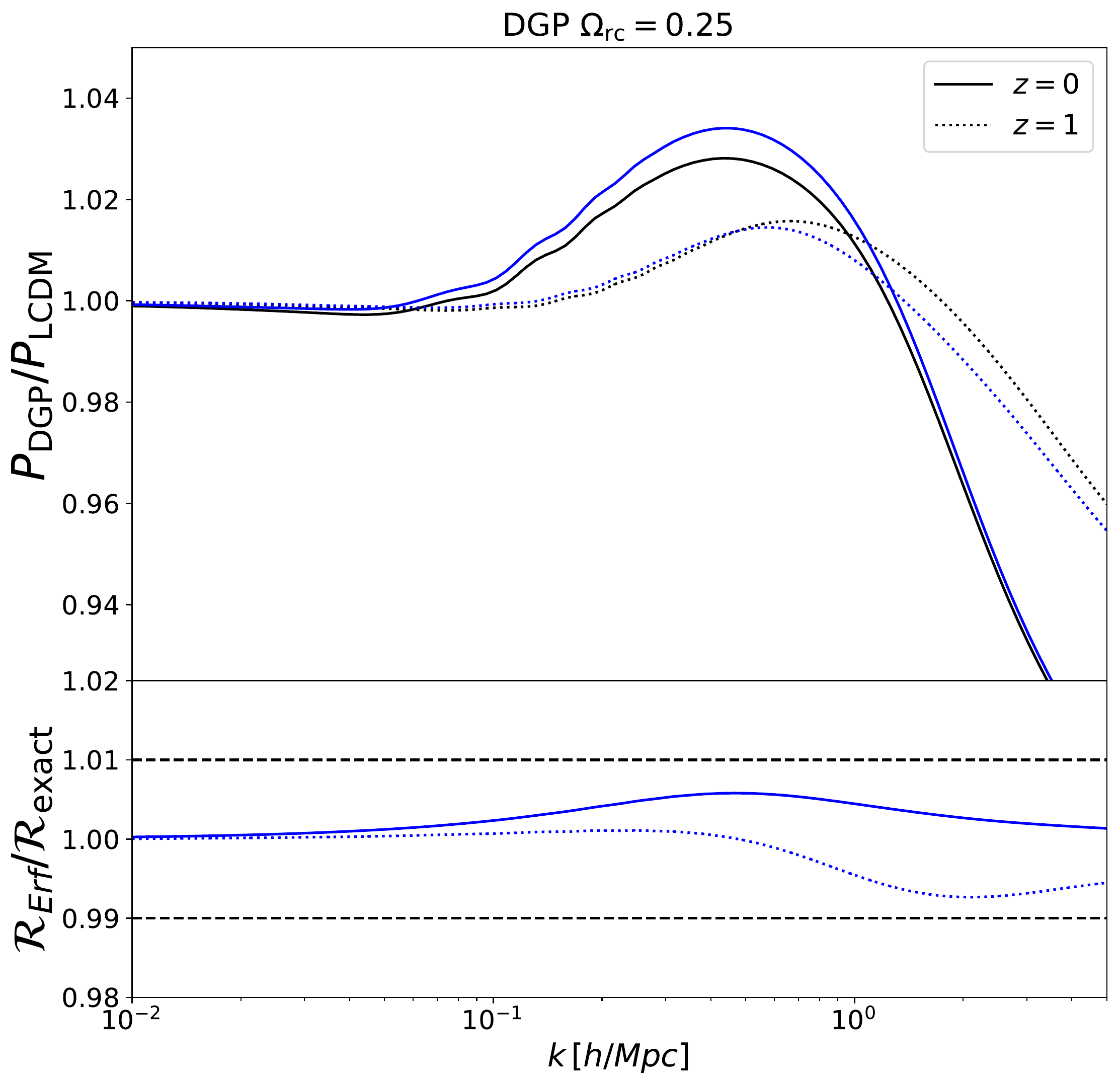}
    \includegraphics[width=0.45\textwidth]{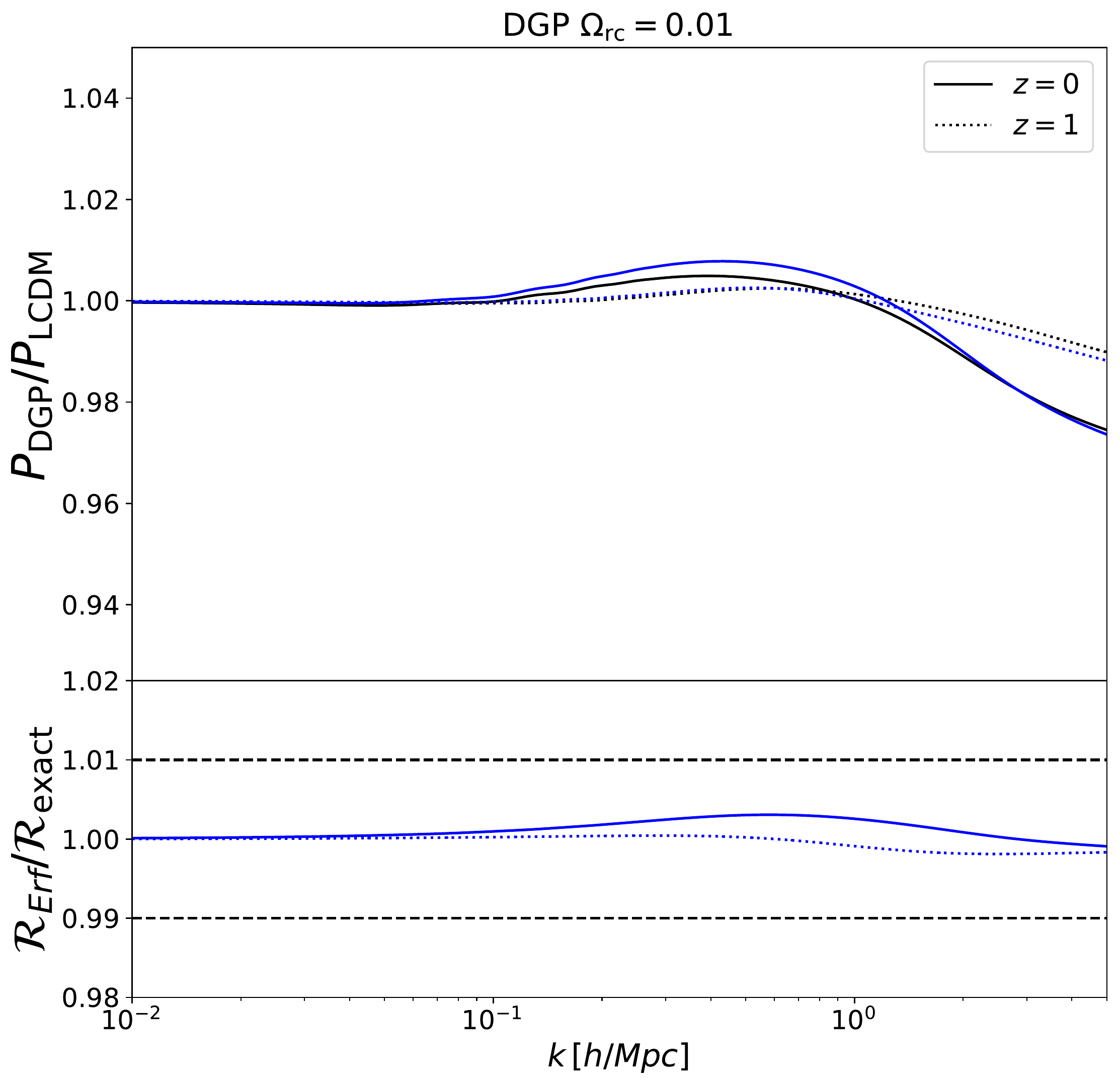}
    \caption{{\bf Top panels:} The ratio of the DGP nonlinear power spectrum to the $\Lambda$CDM nonlinear power spectrum computed using halofit and the halo model reaction (see \autoref{eq:nonlinpk}) for the exact ({\bf black}) and Erf ({\bf blue}) cases. We do not show the nPPF case as it reduces to the exact solution for specific choices of its parameters. The Erf model assumes $\gamma_2=\gamma_3=0$. We have normalised the ratio to unity at large scales for easier comparisons. {\bf Bottom panels:} The ratio of halo model reactions; the Erf model $\mathcal{R}_{\rm Erf}$ to the exact solution. This is equivalent to the ratio of the top panel blue to black curves. We show these results for a moderate modification, $\Omega_{\rm rc}=0.25$ ({\bf left}) and a low modification, $\Omega_{\rm rc} = 0.01$ ({\bf right}). We plot the ratio for two observationally relevant redshifts, $z=0$ ({\bf solid}) and $z=1$ ({\bf dotted}). }
    \label{fig:dgp_Rtest}
\end{figure*}

In the top panels of \autoref{fig:dgp_Rtest} we show the ratio of a DGP power spectrum to a $\Lambda$CDM spectrum with the same background  expansion history, normalised to unity at linear scales for easier comparisons of nonlinear effects. The DGP spectrum is given by \autoref{eq:nonlinpk}. We see the moderate modification gives up to a  $6\%$ deviation from $\Lambda$CDM (above the linear growth enhancement) for $k\leq 3h/{\rm Mpc}$ while the small modification can reach $2\%$ over the same range of scales. Reassuringly, in the bottom panels we find sub-percent agreement between the Erf and exact predictions down to $k=5~h/{\rm Mpc}$, with a smaller disagreement for the smaller deviation from $\Lambda$CDM.

One can further parameterise the time dependence of $q_1$ which would alleviate some of these deviations, but we find these differences to be more than acceptable given the relative size compared to the modification to $\Lambda$CDM shown in the top panels. Moreover, a large number of additional degrees of freedom will be introduced in real data analyses such as intrinsic alignments and parameterisations of baryonic physics. These will be degenerate to some level with modified gravity effects \citep[see][for example]{Schneider:2019xpf}, allowing lower accuracy demands in the modelling of $\mathcal{R}$.

This additional time dependence is highlighted in  \autoref{fig:dgp_Ftest}, where we find that the Erf model can match the exact form of $\mathcal{F}$ extremely well at a fixed redshift.
Upon investigation, we found this dependence to be highly degenerate with $q_2$ which prompted us to not introduce new freedom to the model, especially because we can achieve very good fits already, even without $q_2$.  

Note we have not compared the parametrised model to an emulator nor simulations in this case. Given the excellent agreement with the exact solution we can infer its accuracy is at least as good as the exact solution, given it employs 3 additional degrees of freedom. We remind the reader that the exact reaction was found to be $2\%$ accurate when compared to $N$-body simulations in \cite{Cataneo:2018cic}.
\begin{figure}
    \centering
    \includegraphics[width=0.45\textwidth]{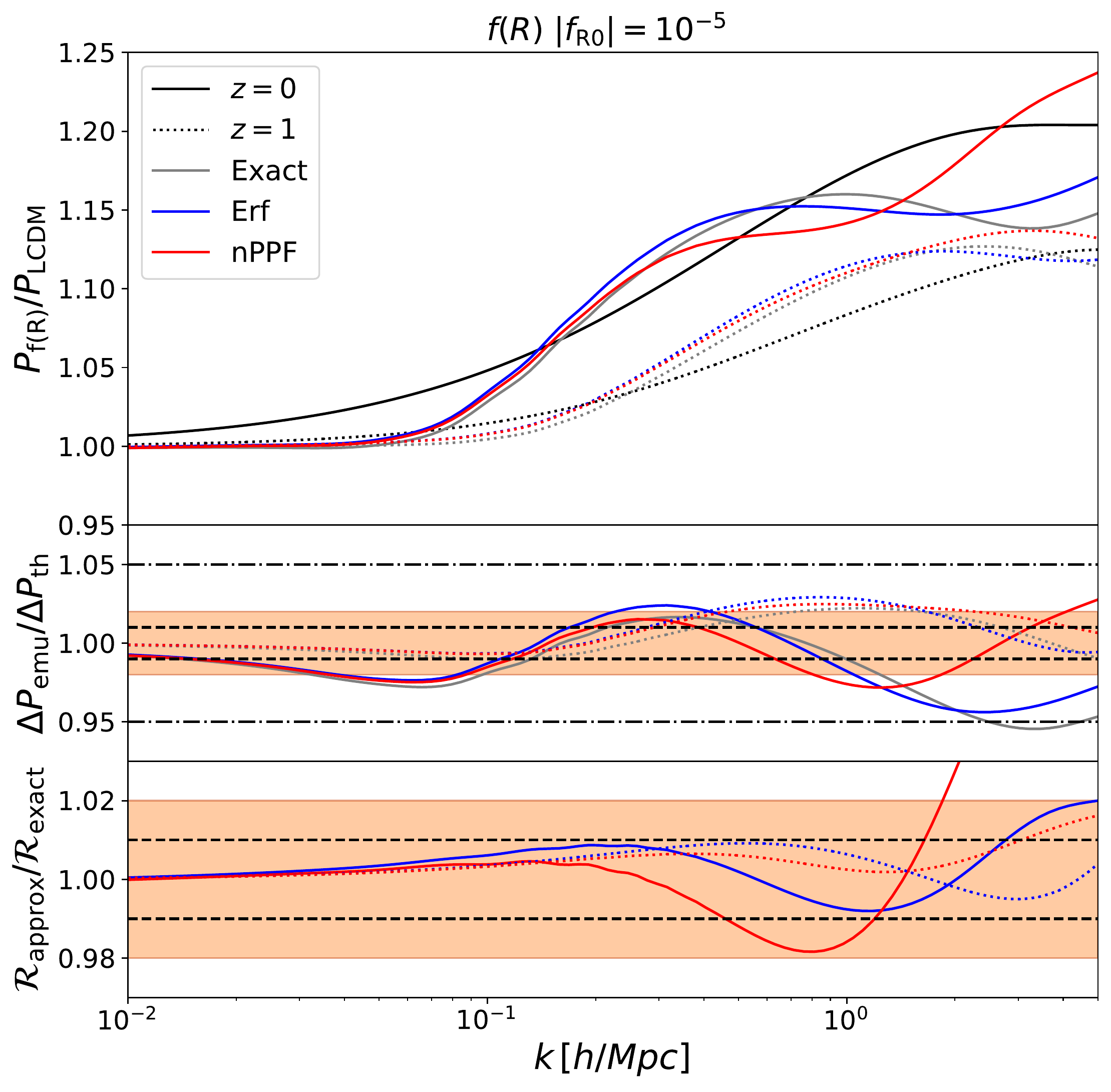}
    \caption{{\bf Top panel:} The ratio of the $f(R)$ nonlinear power spectrum to the $\Lambda$CDM nonlinear power spectrum computed using {\tt HMCode2020} and the halo model reaction (see \autoref{eq:nonlinpk}) for the exact ({\bf grey}), Erf ({\bf blue}) and nPPF ({\bf red}) cases. The fofr emulator is also shown in {\bf black}. {\bf Middle panel:} The ratio between theoretical and emulator predictions for the ratio between $f(R)$ to $\Lambda$CDM spectra, i.e., the ratio of grey, blue and red to black top panels curves. {\bf Bottom panel:} The ratio of halo model reactions; the parameterised models to the exact solution. This is equivalent to the ratio of the top panel coloured curves to the gray curves. Note that both parameterised models have $\gamma_2=\gamma_3=0$. We plot the ratio for two observationally relevant redshifts, $z=0$ ({\bf solid}) and $z=1$ ({\bf dotted}). We show these results for a moderate modification, $|f_{\rm R0}| = 10^{-5}$. The orange bands indicate the $2\%$ region which is the current absolute accuracy of the exact $\mathcal{R}$.}
    \label{fig:fR_Rtest}
\end{figure}
\begin{figure}
    \centering
    \includegraphics[width=0.45\textwidth]{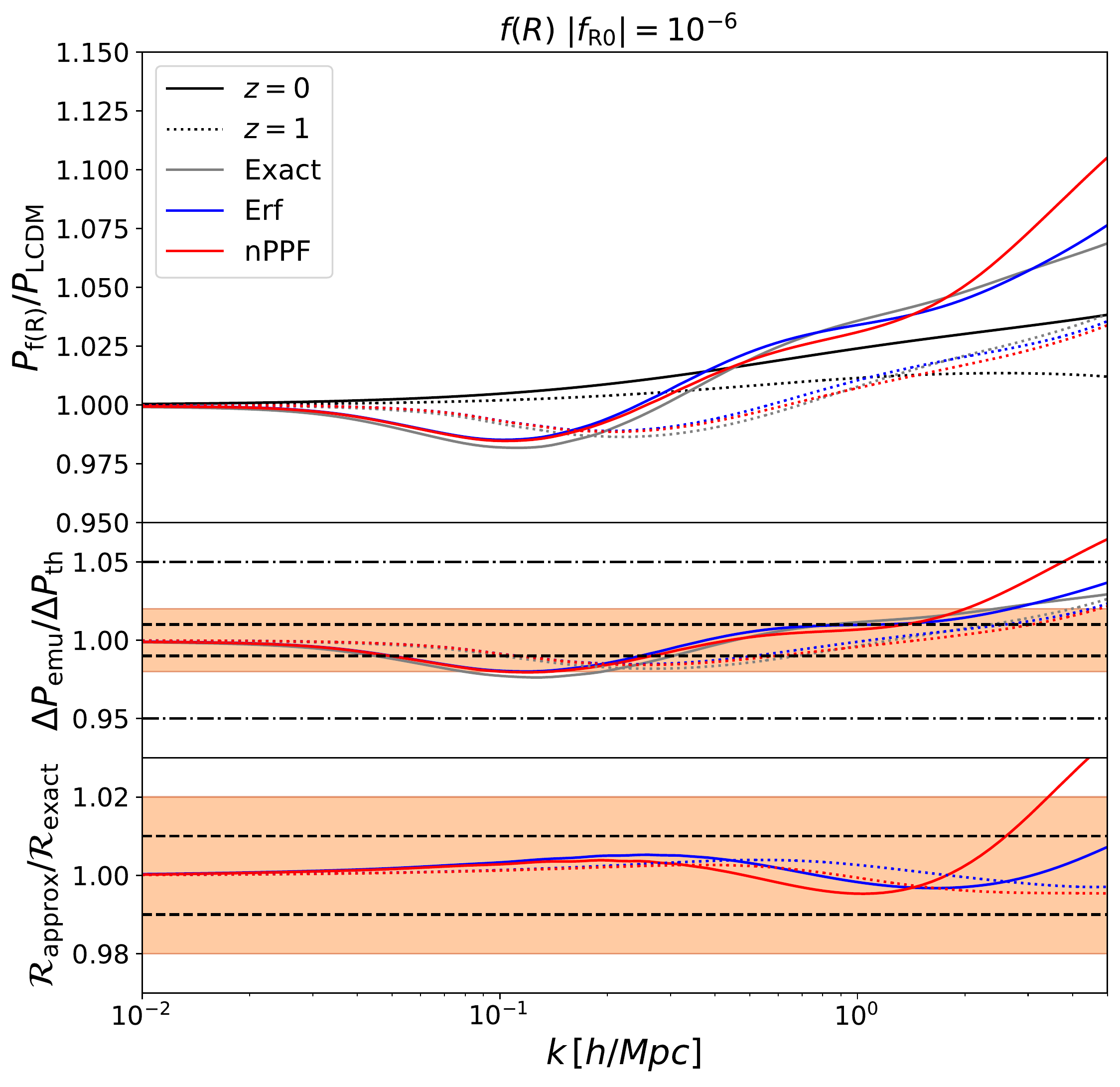}\\
    \includegraphics[width=0.45\textwidth]{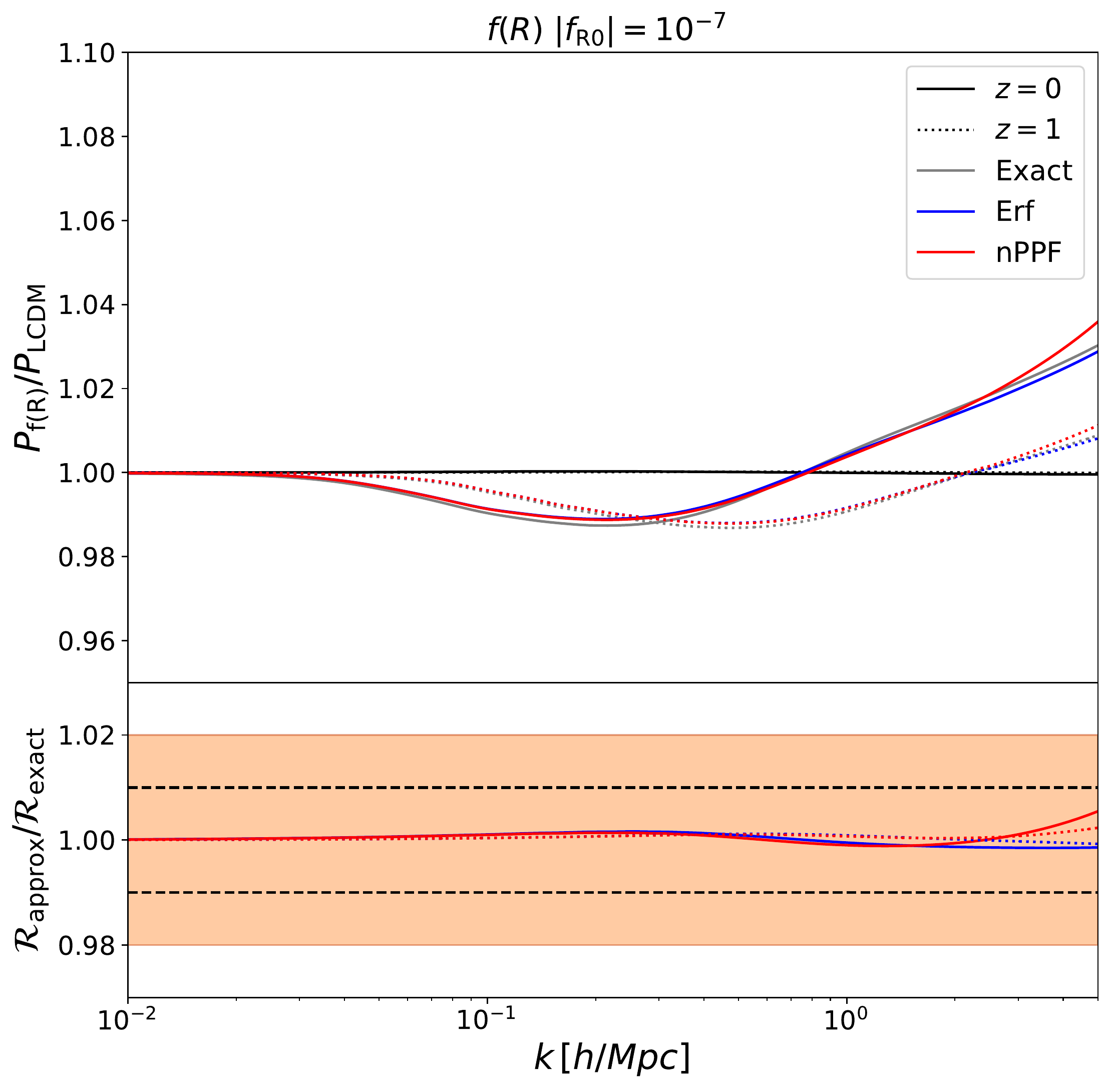}
    \caption{Same as \autoref{fig:fR_Rtest} for a low modification, $|f_{\rm R0}| = 10^{-6}$ ({\bf top}) and a very low modification, $|f_{\rm R0}| = 10^{-7}$ ({\bf bottom}). Note that the fofr emulator for $|f_{\rm R0}| = 10^{-7}$ gives the $\Lambda$CDM prediction and so we omit the middle panel.}
    \label{fig:fR_Rtest2}
\end{figure}


\subsection{Chameleon example: Hu-Sawicki $f(R)$}  \label{sec:valfr} 
For this theory we consider both the nPPF and Erf models for $\mathcal{R}$, and compare them to the exact solution (\autoref{eq:nlexactfr}) as well as at the power spectrum level to the fofr emulator of \cite{Winther:2019mus}. This model makes use of the chameleon screening mechanism which exhibits an environmental and  mass dependence. It also has a Yukawa suppression which returns it to GR at large scales. The additional degree of freedom is the value of the background scalar field at $z=0$, $f_{\rm R0}$, which controls the level of deviation from GR. We consider three levels of deviation from $\Lambda$CDM, $|f_{\rm R0}|=10^{-5}$ (moderate modification), $|f_{\rm R0}|=10^{-6}$ (low modification) and $|f_{\rm R0}|=10^{-7}$ (very low modification). We note that the moderate $f(R)$ modification is already ruled out by data \citep[see][for example]{Cataneo:2014kaa,Desmond:2020gzn,Lombriser:2014dua,Brax:2021wcv}, but provides a good flexibility test of the parameterisation. 

In the nPPF case, we choose the theoretically motivated parameters given in \autoref{eq:frpi}. These emerge from a parameterised form of $f(R)$ gravity \citep{Lombriser:2013eza} and so are approximate. $p_1$ and our new parameter $p_8$ remain free. Treating them both as constants, we fit them in the same way that we fit the Erf model's parameters, by minimising \autoref{eq:chi2}. We note that the other nPPF parameters, $p_2-p_7$, take on different forms for the chameleon screening and Yukawa suppression regimes. We only consider the screening regime which is more relevant for the spherical collapse calculation, and rely on $\mu(\hat{k},a)$ appearing in \autoref{eq:nPPFF} to take care of the Yukawa suppression. 

Yukawa suppression is relevant for large masses, large $y_{\rm env}$ or small values of $f_{\rm R0}$. Given this, we do not expect $p_8$ or $q_4$ to be relevant for spherical collapse where $y_{\rm h} \leq y_{\rm env}\leq 1$, and even less for the 1-halo spectrum where the Sheth-Torman mass function down-weights large masses \citep[see, for example,][]{Schmidt:2008tn}. We verify this by performing two separate fits:  the first only including the parameter sets $\{p_1\}$ and $\{q_1,q_2,q_3\}$ for the nPPF and Erf model respectively, while the second extending these sets to include $p_8$ and $q_4$ respectively. 

We find that values of $q_4,p_8 \geq 0$ negligibly change the goodness of fit for the low and very low modification strengths, while sufficiently negative values degrade the fit, which is expected as the Yukawa scale begins to overlap with the screening scale. Further, we observe only a marginal improvement at $z=0$ for $|f_{\rm R0}|=10^{-5}$ in the Erf case. Given this, all fits shown and quoted here set $q_4=0$ for the Erf case. In the nPPF case, we observe a moderate improvement for $|f_{\rm R0}|=10^{-5}$ and so keep $p_8$. We report the best-fit parameters in \autoref{tab:paramfits}. 

The $f(R)$ results are shown in \autoref{fig:fR_Rtest} and \autoref{fig:fR_Rtest2}. We see the moderate modification can reach a $20\%$ deviation from $\Lambda$CDM for $k\leq 3h/{\rm Mpc}$ while the low and very low modifications reach $10\%$  and $3\%$ respectively. Both parameterisations do well in modelling the moderate modification case $|f_{\rm R0}|=10^{-5}$, shown in \autoref{fig:fR_Rtest}. The Erf model prediction for $\mathcal{R}$ stays within $1\%$ of the exact solution for $k\leq 3h/{\rm Mpc}$. Similarly, the nPPF remains within $2\%$ for $k\leq 2h/{\rm Mpc}$. The situation improves for the lower modification cases, shown in \autoref{fig:fR_Rtest2}. These comparisons exhibit sub-$1\%$ agreement between the Erf (nPPF) model and exact solution for $k\leq 5(3)~h/{\rm Mpc}$ at $z=0$ and $z=1$. 

All power spectra predictions are $\sim 3\%$ consistent with the fofr emulator which mainly demonstrates the accuracy of {\tt HMCode2020}. Interestingly, we find that the additional degrees of freedom within the nPPF and Erf models are degenerate with possible inaccuracies in the pseudo, even down to $k=5~h/{\rm Mpc}$. Again, we leave it to a future work to see if these additional degrees of freedom can improve constraining power on cosmological and gravitational parameters while remaining unbiased.

Our comparisons indicate that for the Erf model, degeneracies between $q_{1-3}$ and $q_4$ make the latter parameter unnecessary. We note that the fit of $q_{1-3}$ becomes insensitive to the value of $q_4$ if it is sufficiently large, here found to be $q_4=0$. For the nPPF model, the additional freedom provided by $p_8$ is necessary to improve the fit, but it does not help substantially for observationally viable values of $f_{\rm R0}$. Further, we remind the reader that we do not know $p_{2-7}$ a priori for unspecified theories of gravity, and so the importance of $p_8$ is likely to be minimal when considering these additional degrees of freedom. 

Lastly, we remark that the Erf model gives a good fit for a range of values for $q_{1-3}$\footnote{Similar fits were found for $\mathcal{O}(0.1)$ values for these parameters.}. The values quoted in \autoref{tab:paramfits} are only the best fit values, which are also very dependent on \autoref{eq:gaussianerr}.  This makes it hard to extract any further dependence on $f_{\rm R0}$ in \autoref{eq:ErfF} (note this already depends on $\mu(k,a)$), which is also beyond the scope of this parametrisation which aims to be general in terms of gravitational degrees of freedom.


\section{Summary}\label{sec:summary}

In this paper we have presented a significant extension of the code described in \cite{Bose:2020wch} which produces nonlinear corrections to the matter power spectrum coming from beyond-$\Lambda$CDM physics in the form of the halo model reaction $\mathcal{R}$. In particular, we have focused on implementing parameterisations of key equations, in particular the background expansion history and the linear and nonlinear Poisson equations.

For the linear scales and background we have considered the effective field theory of dark energy (EFTofDE) while for the nonlinear scales we have considered two distinct parameterisations, a nonlinear parameterised post-Friedmannian (nPPF) based model and a more phenomenological model based on the error function (Erf). Together, these give a general parameterisation of the nonlinear matter power spectrum in Horndeski models. We neglect loop corrections in these parameterisations but leave these as viable additions and we provide theoretical and numerical means of deriving these for the Horndeski class of theories. This being said, we remark that the nonlinear parametrisations are completely general, and so to move beyond the Horndeski class it is sufficient to parametrise only the background expansion history and the linear modification to the Poisson equation. Further, the nonlinear parametrisations also have unscreened limits, and so we are not restricted to theories exhibiting screening. In summary, this work presents a fast, accurate and highly general nonlinear power spectrum predictor for non-standard models of gravity and cosmology including massive neutrinos, parameterised with a minimal set of free, physically meaningful constants.

We have tested these parameterisations against the full solutions for $\mathcal{R}$ in three beyond-$\Lambda$CDM models, wCDM, Hu-Sawicki $f(R)$ and DGP gravity. This has identified a minimal set of 3 free functions of time and 3 dimensionless, positive, $\mathcal{O}(1)$ dimensionless constants, which can replicate the exact solutions to within $1\%$ at $k\leq 5 h/{\rm Mpc}$ and at $z\leq1$ for modifications to GR within current data constraints and within the Horndeski class. This level of imprecision is sub-dominant to the $2\%$ accuracy currently achieved by the reaction method at these scales \citep{Cataneo:2018cic,Cataneo:2019fjp}, and further to the inaccuracies in current pseudo spectrum prescriptions \citep{Bose:2021mkz,Carrilho:2021rqo}. We have seen that the additional parameters have some degree of degeneracy with pseudo spectrum inaccuracies, which may improve the scales of validity for the nonlinear power spectrum $P_{\rm NL}$ as predicted within the halo model reaction framework. We thus suspect that this minimal parametrisation is acceptable for upcoming Stage IV cosmic shear analyses given the flexibility of the nonlinear parameterisation and the many other nuisance degrees of freedom entering a real data analyses, such as those characterising baryonic physics or intrinsic galaxy alignments \citep[see, for example,][]{KiDS:2020ghu}. 

The Erf model is also highly model independent, capturing the basic phenomenology of screening mechanisms. It can thus be suitable for analyses targeting general deviations from $\Lambda$CDM. For example, one may perform a model independent analysis combining the Erf parametrisation with the linear theory growth index $\gamma$-parametrisation \citep{Peebles1980,Linder:2007hg} \citep[also see Eq.~47 of][]{Kennedy:2018gtx} and say the background parametrisation of \cite{Chevallier:2000qy,Linder:2002et}, giving 6 free constants characteristing general deviations from $\Lambda$CDM in the matter power spectrum at a wide range of scales. On the other hand, the nPPF approach is complementary as it can be directly related to specific actions of Nature, making it very suitable when we look to constrain more specific classes of theories. 

In future work we will test the robustness of the minimal parameterisation, and forecast constraints on deviations to $\Lambda$CDM by performing full Markov chain Monte Carlo (MCMC) analyses on mock data of the cosmic shear spectrum. Consistency and accuracy checks can also be performed using recently developed parametrised modified gravity simulations \citep{Hassani:2020rxd,Srinivasan:2021gib,Fiorini:2021dzs,Wright:2022krq,Brando:2022gvg}. On this note, our code is as fast as the original {\tt ReACT} and so is capable of running MCMC analyses. Despite its appreciable baseline speed, we aim to make this even faster by creating emulators based off halo model reaction predictions using the recently released {\tt CosmoPower} code \citep{SpurioMancini:2021ppk} which will highly optimise such analyses. It is a future plan to also perform real data analyses on currently available cosmic shear data to constrain deviations to $\Lambda$CDM using the general minimal parametrisation given in \autoref{tab:parameterisations}. 

It is currently an ongoing project to also extend the halo model reaction to redshift space and biased tracers in a vein similar to \cite{Bose:2019yjp}. We also plan to include interacting dark energy parametrisations \citep{Gleyzes:2015pma,Skordis:2015yra}, a scenario where essentially one decouples the baryons from $\Lambda$CDM modifications, contrary to the scenario considered in this paper where all matter is coupled to the scalar field.

\section*{Acknowledgments}
\noindent The authors would like to thank Matteo Cataneo, Daniel B Thomas, Tessa Baker and Filippo Vernizzi for useful comments and suggestions. They further thank the referee for their useful comments. Lastly, they thank all fautors of \href{https://fautor.org/papers/0005}{fautor.org/papers/0005}. BB, JK and LL acknowledge support from the Swiss National Science Foundation (SNSF) Professorship grant Nos.~170547 \& 202671. BB was supported by a UK Research and Innovation Stephen Hawking Fellowship (EP/W005654/1). ANT acknowledges support from a STFC Consolidated Grant. MT's research is supported by a doctoral studentship in the School of Physics and Astronomy, University of Edinburgh. 
AP is a UK Research and Innovation Future Leaders Fellow [grant MR/S016066/1]. For the purpose of open access, the author has applied a Creative Commons Attribution (CC BY) licence to any Author Accepted Manuscript version arising from this submission.


\section*{Data Availability}

The software used in this article is publicly available in the {\tt ACTio-ReACTio} repository at \url{https://github.com/nebblu/ACTio-ReACTio}. In the same repository we also provide two {\tt Mathematica} notebooks: \href{https://github.com/nebblu/ACTio-ReACTio/tree/master/notebooks}{{\tt GtoPT.nb}} that explicitly calculates the modified gravity 1st, 2nd and 3rd order Poisson equation modifications for particular covariant theories of gravity as well as provides maps between the EFTofDE $M$- and $\alpha$-bases, and \href{https://github.com/nebblu/ACTio-ReACTio/tree/master/notebooks}{{\tt Nonlinear.nb}} which provides expressions, tests and comparisons of the nonlinear Poisson equation modifications.


\bibliographystyle{mnras}
\bibliography{mybib}

\appendix 

\section{Linear Poisson modifications} \label{sec:efttopoisson}

\subsection{Horndeski in the $\alpha$-basis } \label{app:alphamu}
The linear modification to the Poisson equation in the Horndeski class of EFTofDE, under the quasi-static approximation, and in the $\alpha$-basis, $\{ H , \alpha_M, \alpha_B, \alpha_K, \alpha_T \}$, is given by 
\begin{equation}
\mu (k,a) = \frac{2}{\kappa^2} \frac{f_1(a) + f_2(a) a^2/k^2}{f_3(a) + f_4(a) a^2/k^2} \, , \label{eq:mueft}
\end{equation} 
where the constituent functions are given by 
\begin{align}
    f_1 = &  B_2 C_3 - C_1B_3 \, , \\ 
    f_2 = &  B_2 C_\pi \, , \\ 
    f_3 = &  A_1[B_3 C_2 - B_1C_3(k,a)] \nonumber \\ 
            & + A_2[B_1 C_1 - B_2 C_2] \nonumber \\ 
            & + A_3[B_2C_3 - B_3 C_1] \, , \\ 
    f_4 = &  [A_3 B_2 - A_1B_1]C_\pi \, . 
\end{align}
Finally, we give the $A$, $B$ and $C$ functions in terms of the $\alpha$-basis as \citep{Pogosian:2016pwr}
\begin{align}
    A_1 = &  2 M^2 \, , \\ 
    A_2 = & \alpha_B H M^2 \, , \\
    A_3 = & 0 \, , \\ 
    B_1 = & -\frac{1}{c_T^2} \, , \\ 
    B_2 = & 1 \, , \\ 
    B_3 = & \frac{(-\alpha_M + \alpha_T) H}{c_T^2} \, , \\ 
    C_1 = & - B_3 M^2 c_T^2 \, , \\ 
    C_2 = & \frac{A_2}{2}  \, , \\ 
    C_3 = & c + \frac{H M^2}{2} \Big[-2 \alpha_T H + \alpha_M^2 c_T^2 H + a H\alpha_B' + a H \alpha_M' \nonumber \\ 
    & + a \alpha_T H \alpha_M' + 2 a \alpha_T H'  + a^2 \alpha_T' H' + \alpha_B[(1+\alpha_M)H + a H'] \nonumber \\ 
    & +\alpha_M[H(1-\alpha_T + 2 a \alpha_T') + ac_T^2 H'] + a^2 H \alpha_T''\Big] \, , \\ 
    C_\pi = & -\frac{1}{4} a H \Big[12 c H' + H M^2 (6 \alpha_M^2 c_T^2 H H' \nonumber \\ 
     & + 6 \alpha_B(2 a (H')^2 +H [(4 + \alpha_M)H' + a H''] )  \nonumber \\ 
     & + \alpha_M(c_T^2(12 a(H')^2 - \bar{R}')+6 H (2(2c_T^2 + a \alpha_T') H' + a c_T^2H'')) \nonumber \\ 
     & + a[12 (\alpha_T + a \alpha_T') (H')^2  \nonumber \\ 
     & - \alpha_T' \bar{R}' + 6 H (H'(\alpha_B'+c_T^2\alpha_M' + 5 \alpha_T' + a \alpha_T'') + a \alpha_T' H'')]) \Big] \, , 
\end{align}
where $c_T^2 = (1+\alpha_T)$ and $\bar{R}$ is the background Ricci scalar. The $c(a)$ parameter (\autoref{eq:cmbasis}) in the $\alpha$-basis is given by 
\begin{align}
    c(a) & = -M^2 \Big[ \frac{3 H_0^2 \Omega_{m,0}}{2 a^3 \kappa^2 M^2}  \nonumber \\ 
      & + \frac{1}{2} H \Big(a H' [ (2+\alpha_M) c_T^2 + a \alpha_T']  \nonumber \\
      & + H( c_T^2[(\alpha_M-1)\alpha_M + a \alpha_M'] + 2 a \alpha_M \alpha_T' + a^2 \alpha_T'')\Big)  \Big] \, ,\label{eq:calpha}
\end{align}
where we have used $\rho_m = 3 H_0^2 \Omega_{m,0}/(\kappa^2 a^3)$, with $\Omega_{m,0}$ being the matter density fraction today and $H_0 = H(a=1)$ is the Hubble constant.

We note that in our code we make the redefinition $M^2 \kappa^2 =  M^2/m_0^2 \rightarrow M^2 $ where $m_0^2($ is the Planck mass.  Further, we comment on the flexibility offered here. One can choose to specify any two of $\{ \alpha_M, M^2, H\}$. If $H$ is specified then either $\alpha_M$ or $M^2$ must also be specified, with the third function given by the relation in \autoref{eq:am}. If $H$ is not specified, then we must solve the Friedmann equations to obtain $H$. As a default in our code, $H$ is specified and it is assumed that the specified expressions for $\alpha_M$ and $M^2$ are consistent with \autoref{eq:am}.

We can also take the small scale ($k\rightarrow \infty$) limit of \autoref{eq:mueft} to get a simpler expression valid at scales where the QS is a safer approximation and for models exhibiting negligible scale dependence in the linear growth. This is given by 
\begin{equation}
    \mu_\infty = \frac{1}{M^2 \kappa^2} \Big(1+\alpha_T + \beta_\xi\Big) \, , \label{eq:mukinf}
\end{equation}
where 
\begin{equation}
    \beta_\xi = \frac{2}{c_s^2 \alpha} \Big(c_T^2 \frac{\alpha_B}{2} + \alpha_M - \alpha_T \Big)^2 \, 
\end{equation}
with $c_s^2$ and $\alpha$ and $\alpha$ given by \autoref{eq:cs2} and \autoref{eq:alpha}.

\subsection{Example: Hu-Sawicki $f(R)$} \label{app:fR}
In this section we derive the relevant EFTofDE parameters and linear Poisson modification for the Hu-Sawicki form of $f(R)$ gravity \citep{Hu:2007nk}. This exercise is also performed in the {\tt GtoPT} notebook provided in the {\tt ACTio-ReACTio} github repository. 

The action in $f(R)$ gravity is given by  
\begin{align}
    S=& \int d^4x \sqrt{-g} \frac{1}{2\kappa^2}  \Big( R+f(R) \Big) \nonumber \\
    \approx & \int d^4x \sqrt{-g} \frac{1}{2\kappa^2}  \Big( R+f(\bar{R}) + f_R(\bar{R}) (R-\bar{R}) \Big) \nonumber\\
    =& \int d^4x \sqrt{-g} \frac{1}{2\kappa^2}  \Big( (1+f_R)R+f- f_R\bar{R} \Big) \, ,
\end{align}
where $f_R = df(R)/dR$ and we have performed a Taylor expansion in the second line. We can then map this action onto the functions given in \autoref{eq:horndeski} together with an identification of the scalar degree of freedom  $\phi \equiv (1+f_R)/\kappa^2$ \citep{deFelice2011_linperturb}:
\begin{align}
  G_2 = -\frac{1}{2\kappa^2} (\bar{R}f_R - f), \, \, \, \,  G_3 = 0, \, \, \, \, \mathrm{and} \, \, \, \,   G_4 = \frac{1}{2\kappa^2} (1+ f_R) \, . \label{eq:fRGs}
\end{align} 
If we now write down the action in the ADM formalism and use the Gauss-Codazzi relation, we can compare to \autoref{eq:s01} and \autoref{eq:s2} to get 
\begin{align}
\label{eq:FRinEFT}
    \Omega = (1+f_R), \,  \Lambda = \frac{1}{2\kappa^2} (f-\bar{R}f_R), \,  \ c = \bar{M}^2_2 = \bar{M}_1^3 = M_2^4 = 0  \, .
\end{align}
Using \autoref{eq:am} - \autoref{eq:ak} we have 
\begin{align} 
\alpha_M &= \frac{a f_R'}{1+f_R}  \, , \label{eq:fram} \\ 
\alpha_T &= 0  \, , \label{eq:frat}  \\ 
\alpha_B &= - \frac{a f_R'}{1+f_R} \, , \label{eq:frmb}  \\ 
\alpha_K &= 0 \, , \label{eq:frak}
\end{align} 
with 
\begin{equation}
 M^2  = \frac{(1+f_R)}{\kappa^2} \, , \label{eq:frm2} \\ 
\end{equation}
where a prime denotes a scale factor derivative. When substituting into the expressions in Appendix~\ref{app:alphamu} we get the following solution for $\mu$ (\autoref{eq:mueft})
\begin{equation}
    \mu = \frac{1}{1+f_R}\Big[1 + \Big(\frac{k}{a}\Big)^2 \frac{1}{3 \tilde{\Pi}(k,a)} \Big] \, , \label{eq:muhu1}
\end{equation}
 where 
\begin{equation}
    \tilde{\Pi}(k,a) = \left(\frac{k}{a}\right)^2 + (1+f_R) \frac{\bar{R}_f}{3} \, , \label{eq:pitilde} 
\end{equation}
 and 
 \begin{equation}
     \bar{R}_f \equiv \frac{d \bar{R} }{d f_R} = f_{RR}^{-1}  \Big(=  \frac{\bar{R}'}{f_R'} \Big) \, . \label{eq:rbarf}
 \end{equation}
 
In the Hu-Sawicki model we have the following choice for $f(R)$ 
\begin{equation}
f(R)  = -m^2 \frac{c_1 (R/m^2)^n}{c_2(R/m^2)^n+1} \, , 
\label{eq:husawfR}
\end{equation}
where in this work we set the index $n=1$ and the mass scale $m^2$, $c_1$ and $c_2$ are free parameters to be constrained by data. Taking the derivative of \autoref{eq:husawfR} with respect to $R$ and the high curvature limit ($R\gg m^2$) gives 
\begin{equation}
    f_R = - \frac{c_1}{c_2^2} \Big(\frac{m^2}{R}\Big)^2 \, .  \label{eq:fRform}
\end{equation}
By rearranging this equation and evaluating at the background level at $a=1$ (today), we can apply the following standard reparameterisation
\begin{equation} 
\frac{c_1}{c_2^2} = - \bar{f}_{R0} \Big(\frac{\bar{R}_0}{m^2}\Big)^2  \, ,  \label{eq:c12}
\end{equation}
where $\bar{f}_{R0}$ is the background value of $f_R$ evaluated today and is a free parameter governing the level of modification to $\Lambda$CDM at the level of structure formation. Substituting \autoref{eq:c12} into \autoref{eq:fRform} gives
\begin{equation} 
f_R = \bar{f}_{R0} \Big(\frac{\bar{R}_0}{R}\Big)^2 \, . 
\end{equation}
Further, we have 
\begin{equation} 
f_{RR}= -2 \bar{f}_{R0} \Big(\frac{\bar{R}_0}{R}\Big)^2  \frac{1}{R}\, . 
\end{equation}
Using the background expression for $f_{RR}$ in \autoref{eq:rbarf} and substituting into \autoref{eq:pitilde} gives 
\begin{equation}
\tilde{\Pi}(k,a) = \left(\frac{k}{a}\right)^2 + (1+f_R) \frac{1}{6 |f_{R0}|}\frac{\bar{R}^3}{\bar{R}_0^2 } \, . \label{eq:ptilde2}
\end{equation}

Now if we approximate the background to be close to $\Lambda$CDM, as supported by observations and by construction for $|f_{\rm R0}|\ll1$, we have 
\begin{equation}
\bar{R} \approx 3 \frac{H_0^2}{a^3} \Big( \Omega_{m,0} + 4 a^3 \Omega_{\Lambda,0} \Big) \, ,
\end{equation}
where $\Omega_{\Lambda,0} = 1 - \Omega_{m,0}$ for a flat $\Lambda$CDM universe. Taking $a=1$ we have the curvature today 
\begin{equation}
\bar{R}_0 \approx 3 H_0^2 \Big( 4 - 3\Omega_{m,0} \Big) \, .
\end{equation}
Finally substituting $\bar{R}$ and $\bar{R}_0$ in \autoref{eq:ptilde2} we get the expression for $\mu$ as it appears in {\tt ACTio-ReACTio} \citep{Bose:2020wch} 
\begin{equation}
\mu = 1 + \Big(\frac{k}{a}\Big)^2 \frac{1}{3 \Pi(k,a)} \, , \label{eq:muhu2}
\end{equation}
with
\begin{align}
\Pi(k,a) =& \left(\frac{k}{a}\right)^2+\frac{\Xi(a)^3}{2f_0(4 - 3\Omega_{\rm m,0})^2}, \\   \Xi(a) =& \frac{\Omega_{\rm m,0}+4a^3\Omega_{\Lambda,0}}{ a^3},
\end{align}
where $f_0 = |{f}_{\rm R0}|/H_0^2$.

We make the crucial note here that in the derivation above we have over-constrained our system. Namely we have specified all of $\Omega, c$ as well as set $H = H_{\rm \Lambda CDM}$. If we use \autoref{eq:fram}-\autoref{eq:frm2} together with  $H = H_{\rm \Lambda CDM}$ we find that $c \neq 0$ and we don't recover \autoref{eq:muhu2}. This follows directly from the fact that $H_{\rm \Lambda CDM}$ is not an exact solution for the Friedmann equations in $f(R)$ which implicitly assumes $c=0$. 

In our code we give the option to over constrain by specifying $c(a)$. Alternatively, one can code in the Friedmann equations and solve for $H(a)$. This of course increases computational inefficiency. To partially alleviate this issue, we can also place the additional constraint on $c_s^2(a)$ instead of $c$ which has well motivated physical priors (see text around \autoref{eq:cs2}). The relationship between $c_s^2(a)$ and $c(a)$ is derived from \autoref{eq:cs2} and \autoref{eq:calpha}. It is given explicitly in the {\tt GtoPT} notebook as well as the {\tt Actio-Reactio} source code.

The derivation of the 2nd and 3rd order modifications to the Poisson equation (see \autoref{eq:Perturb3}) from the ADM decomposed action requires us to go to higher order in the metric perturbations. We do not do this here as we omit these corrections from our code due to computational difficulty and the low level of impact they have on the final nonlinear power spectrum (see Section~\ref{sec:validation}). However, in the provided Mathematica notebook, {\tt GtoPT}, one can go from a specified $G_i$ of the Horndeski Lagrangian to $\mu$, $\gamma_2$ and $\gamma_3$ following the map given in \cite{Bose:2016qun}. We provide a number of examples in that notebook and refer the reader to \cite{Bose:2020wch,Bose:2016qun} for the forms of $\mu$, $\gamma_2$ and $\gamma_3$ in DGP and Hu-Sawicki $f(R)$ gravity.


\section{Nonlinear Poisson modifications} \label{app:nonlin}

\subsection{Exact forms}
We provide the exact forms for the nonlinear modification to the Poisson equation  (see \autoref{eq:poisson2}) in DGP and $f(R)$ gravity, which are reproduced from \cite{Bose:2020wch}.

The modification in DGP is given by \citep{Schmidt:2009yj}
\begin{equation}
    \mathcal{F}_{\rm DGP} = \frac{2}{3\beta(a)} \frac{\sqrt{1+s^{3}} -1}{s^{3}}, \label{eq:nlexactdgp}
\end{equation}
where\footnote{We note a typo appearing in Eq. C7 of \cite{Bose:2020wch} where $\delta$ should have been $(\delta+1)$.}
\begin{equation}
    s  = \left[ \frac{2 \Omega_{m,0} (\delta+1) }{9 a^3 \beta(a)^2 \Omega_{rc}} \right]^{\frac{1}{3}},
\end{equation}
$\delta$ being the nonlinear over-density given by 
\begin{equation}
    \delta = y^{-3} (1+\delta_i) -1 \, , \label{eq:nldensity} 
\end{equation}
with $\delta_i$ being the initial over-density and 
\begin{equation}
    y \equiv \frac{R_{\rm TH}/a}{R_i/a_i} \, , \label{eq:ydef}
\end{equation}
$R_{\rm TH}$ and $R_i$ being the physical halo top-hat radius at the target scale factor $a$ and the initial scale factor $a_i$ respectively. $\Omega_{\rm rc} \equiv 1/(4 H_0^2 r_c^2)$ where $r_c$ is the cross-over scale and is the free parameter of the theory governing the level of modification. Finally, $\beta(a)$ is given by 
\begin{equation}
\beta(a) \equiv 1+\frac{H}{H_0} \frac{1}{\sqrt{\Omega_{\rm rc}}}   \left(1+\frac{aH'}{3H}\right) \,.
\end{equation}

The fully nonlinear modification in  Hu-Sawicki $f(R)$  is given by \citep{Lombriser:2013eza}
\begin{equation}
    \mathcal{F}_{\rm fR} = {\rm min} \left[O - O^2 + \frac{O^3}{3}  , \frac{1}{3} \right] , \label{eq:nlexactfr}
\end{equation}
where 
\begin{equation}
    O = \frac{ f_0 y_h a (3 \Omega_{m,0} - 4)^2}{ \Omega_{m,0} (R_i/a_i)^2} \times
    \left[ \tilde{G}(y_{\rm env}) -  \tilde{G}(y_{\rm h}) \right] \, ,
\end{equation}
and 
\begin{equation}
    \tilde{G}(y) = \left[ \frac{\Omega_{m,0}}{(y a)^3} +4 - 4\Omega_{m,0} \right]^{-2} \, , 
\end{equation}
where $y_h$ is the quantity solved for using $f(R)$ halos whereas $y_{\rm env}$ is that solved for in the environment, which is approximated by performing the same calculation but with $f_0 = 0$.

\subsection{nPPF forms} \label{app:nppfforms}
We also reproduce the nPPF expressions for both of these theories from \cite{Lombriser:2016zfz}. In DGP we have the following values for the $p_i$ parameters in \autoref{eq:nPPFF}
\begin{align}
    p_1 & = 2, \qquad   p_2 = 1,  \qquad p_3 = \frac{3}{2}, \nonumber \\ 
    p_4(a) &= 2 \Big(\frac{\Omega_{\rm m,0}}{4 \Omega_{\rm rc} } \frac{1}{9 \beta(a)^2} \Big)^{1/3} ,  \qquad p_5  = -1, \qquad  p_6 = 0, \nonumber \\  p_7 & = 0 \, , \label{eq:dgppi}
\end{align}
which reproduce \autoref{eq:nlexactdgp} exactly. Note if using \autoref{eq:nPPFF0} we simply set $p_2 = \frac{1}{3\beta(a)}$. 

On the other hand, the Hu-Sawicki $f(R)$ parameterisation is not exact but is closely matched by the following parameters in the screening regime \citep{Lombriser:2016zfz} for a given choice of $p_1$ (using \autoref{eq:nPPFF})
\begin{align}
  p_2 & = 1,  \qquad p_3 = 7,  \nonumber \\ 
    p_4 &= 2 \Omega_{\rm m,0}^{1/3}\Big[(\Omega_{\rm m,0} + 4(1-\Omega_{\rm m,0})^{-2} \frac{p_1}{3 |f_{\rm R0}|} \Big]^{1/p_3} ,\nonumber  \\ 
     p_5  &= -1, \qquad  p_6 = \frac{2}{3 p_3}, \qquad p_7 = -\frac{6}{7} \, , \label{eq:frpi}
\end{align}
where we used $\alpha= 1/(n+1) = 0.5$ \citep{Lombriser:2013eza} in Equation~5.6 of \cite{Lombriser:2016zfz}. Again, if using \autoref{eq:nPPFF0} we set $p_2=\frac{1}{3}$. 

\subsection{Comparisons}\label{sec:nlfcomparisons}
Here we provide some comparisons of the approximate expressions for $\mathcal{F}$ given by the nPPF model (\autoref{eq:nPPFF}) and the Erf model (\autoref{eq:ErfF}) against the exact expressions in DGP (\autoref{eq:nlexactdgp}) and $f(R)$ (\autoref{eq:nlexactfr}). Since the nPPF form is exact for DGP, we only compare it in the $f(R)$ case. Unless otherwise stated, the fits are performed as described in \autoref{sec:validation} and shown in \autoref{tab:paramfits}.

\subsubsection{DGP}

\begin{figure}
    \centering
    \includegraphics[width=0.45\textwidth]{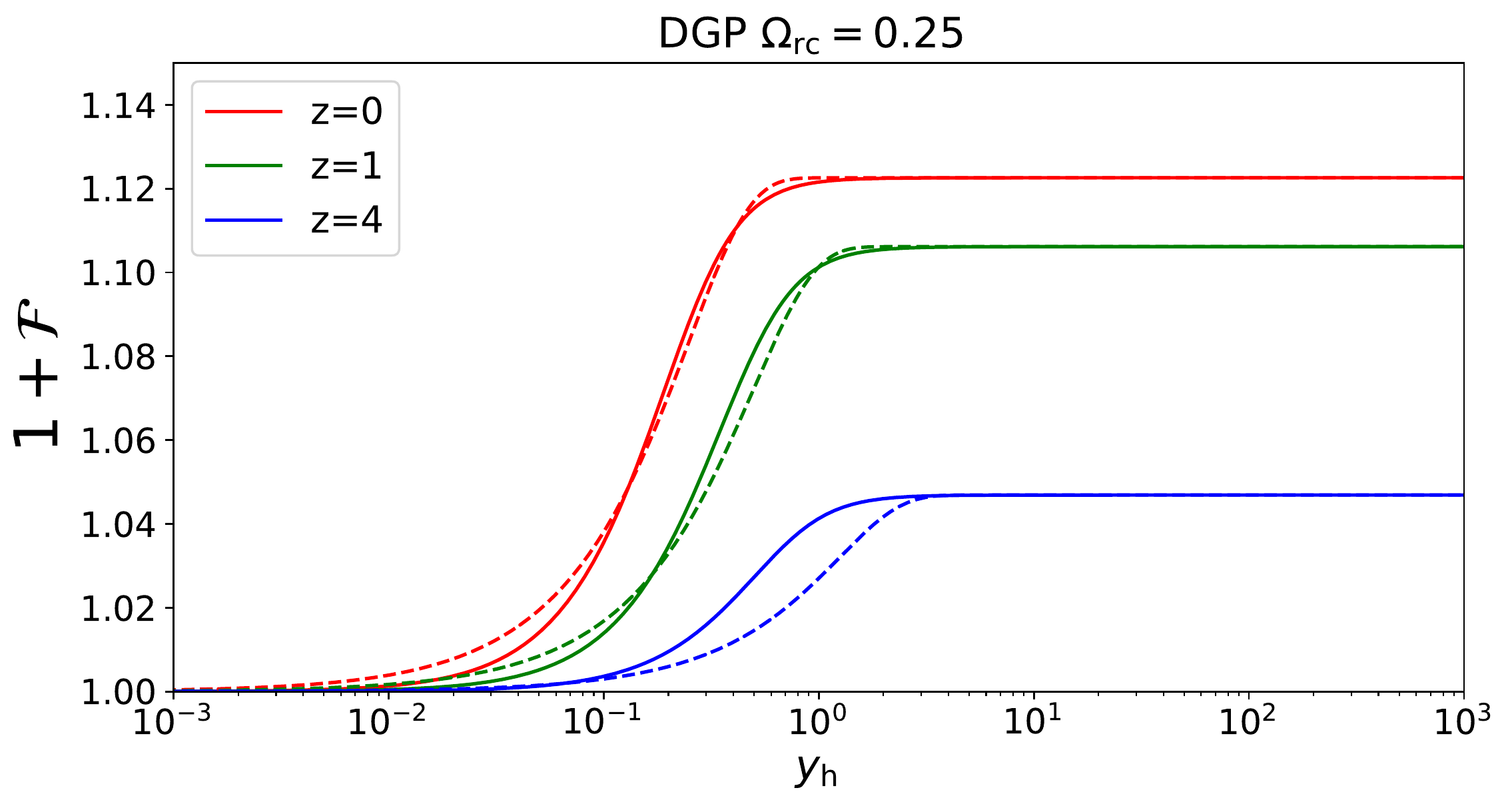}  
    \includegraphics[width=0.45\textwidth]{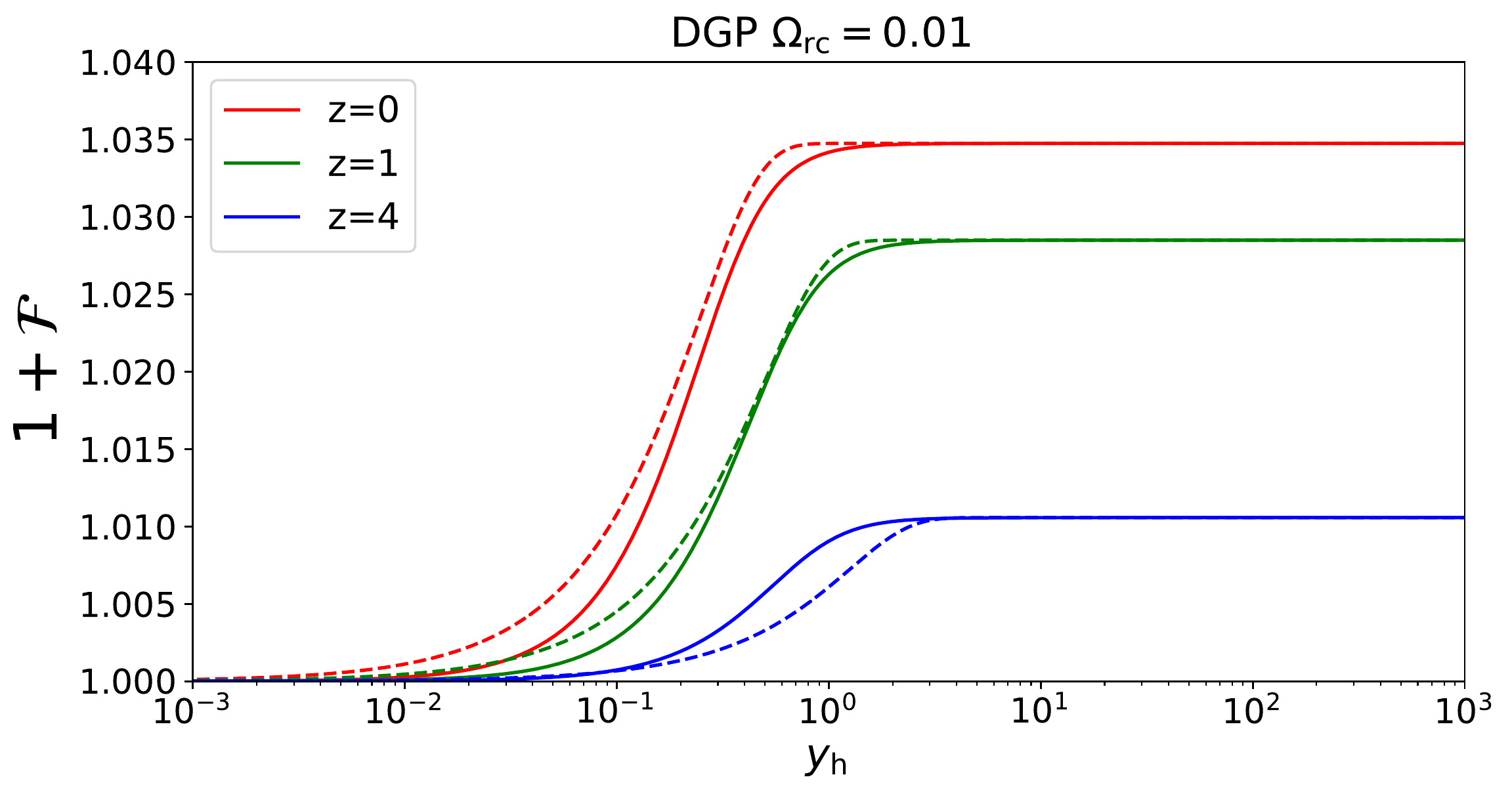}
    \caption{The modification to the Poisson equation $1+\mathcal{F}$ (see \autoref{eq:poisson2}) in DGP gravity for $\Omega_{\rm rc}=0.25$ ({\bf left}) and $\Omega_{\rm rc} = 0.01$ ({\bf right}). We plot the modifications as a function of normalised halo top-hat radius parameter for three different redshifts, $z=0$ ({\bf red}), $z=1$ ({\bf green}) and $z=4$ ({\bf blue}). The {\bf solid} curves are the exact solution while the {\bf dashed} curves are made using a single parameter fit to the exact $\mathcal{R}$ using \autoref{eq:ErfF} (see \autoref{tab:paramfits}).}
    \label{fig:dgp_Ftest}
\end{figure}

In \autoref{fig:dgp_Ftest} we show the nonlinear modification to the Poisson equation, $1+\mathcal{F}$, in DGP for $\Omega_{\rm rc}=0.25$ and $\Omega_{\rm rc}=0.01$ as a function of top-hat radius parameter $y_{\rm h}$. We plot the exact solution given by \autoref{eq:nlexactdgp} given as solid curves to the best fit Erf model given as dashed curves. We see an additional redshift dependence of the screening scale becoming important for high $z$. As modifications to GR are expected to be small at high redshift, this deviation may not be so important, which is supported by \autoref{fig:dgp_Rtest}. We have performed a fit of this redshift dependence and find it behaves very well as a power law, with an $\mathcal{O}(0.1)$ exponent (specifically $\leq 0.15$), which is first of all small and second of all degenerate with $q_2$ and $q_3$, meaning the model likely has sufficient freedom to very well capture a DGP type of modification to gravity without biasing cosmological or gravitational constraints.

\subsubsection{Hu-Sawicki $f(R)$}

Here we check that the nPPF and Erf models can qualitatively reproduce the exact form of $\mathcal{F}$ (\autoref{eq:nlexactfr}) across all scales, masses and environments for Hu-Sawicki $f(R)$ gravity. 

Before showing the results, we make a note on the best fitting parameters for the Erf model. We find that the fits in \autoref{tab:paramfits}, performed by fitting the exact prediction for the reaction $\mathcal{R}$, do not give a very good agreement when comparing to the exact form of $\mathcal{F}$. In particular, we find that the mass dependence parameter, $q_2$, seems to be underestimated when fitting  $\mathcal{R}$. This parameter dictates the left hand slope in the contour plots in this section. Such a discrepancy may be due to a number of factors including a missing redshift dependency, the details of the fit, degeneracies with $q_3$ and failings of the power law description. We find a better by-eye fit for $\mathcal{F}_{\rm Erf}$ across redshifts $z=0,1,4$ and all values of $f_{\rm R0}$ is $q_2 = 0.85$. All other parameters are as in \autoref{tab:paramfits} unless otherwise stated.

\begin{figure*}
    \centering
    \includegraphics[width=0.95\textwidth]{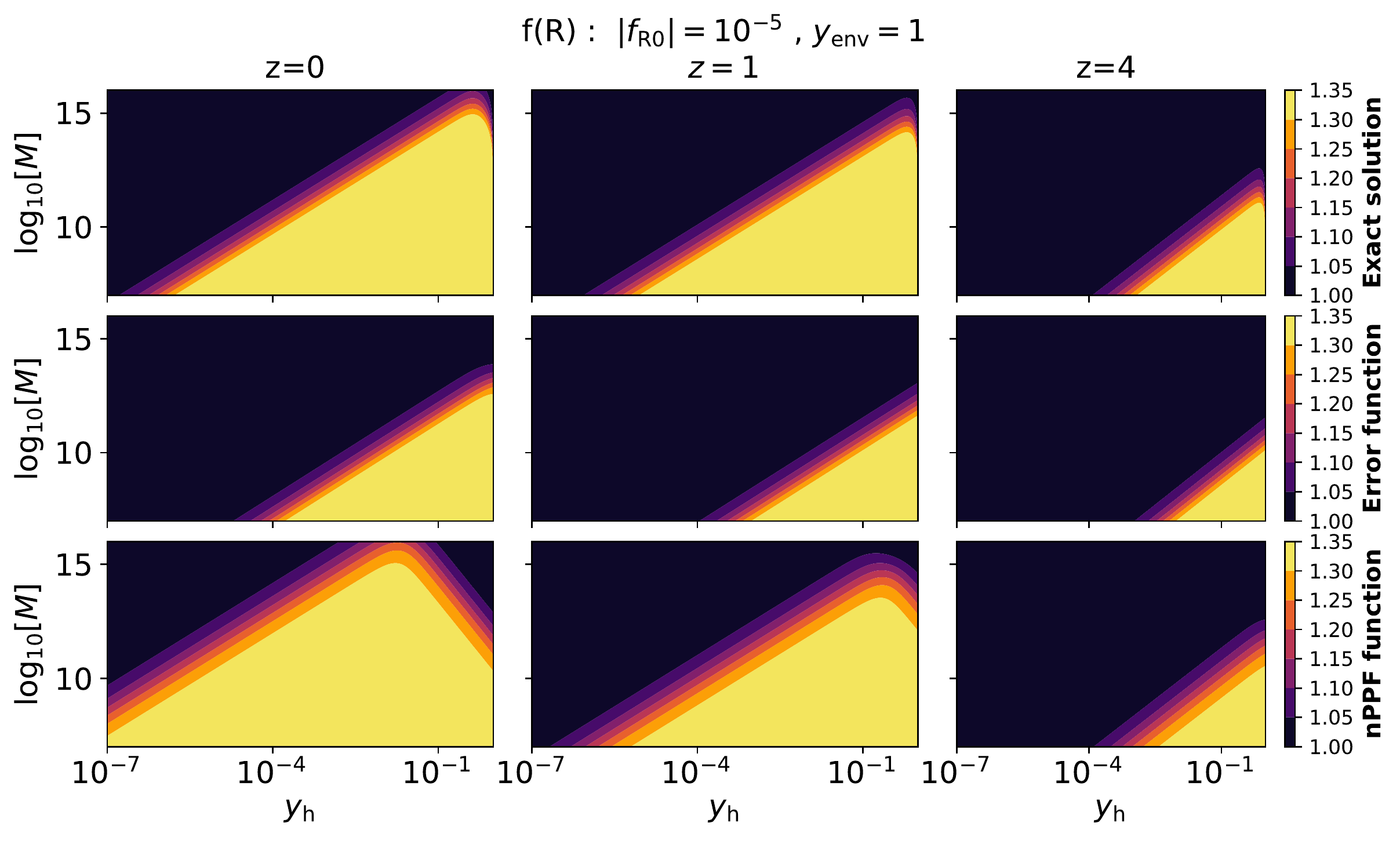}
    \caption{The modification to the Poisson equation $1+\mathcal{F}$ (see \autoref{eq:poisson2}) in Hu-Sawicki $f(R)$ as a function of $\log_{10}(M)$ and top-hat radius parameter $y_{\rm h}$. We set $|f_{\rm R0}| = 10^{-5}$ and $y_{\rm env}=1$. The {\bf top} panels show the exact solution, the {\bf middle} panels show the phenomenological solution based on the error function and the {\bf bottom} panels show the nPPF function. The {\bf left} most column shows the functions for $z=0$, the {\bf middle} for $z=1$ and the {\bf right} most column for $z=4$.}
    \label{fig:f5_Ftest_yenv1}
\end{figure*}

In \autoref{fig:f5_Ftest_yenv1} we show $1+\mathcal{F}$ for the exact (top panels), the Erf (middle panels) and the nPPF (lower panels) cases with $|f_{\rm R0}| = 10^{-5}$, characterising a moderate modification to $\Lambda$CDM. We do not show the $|f_{\rm R0}| = 10^{-6},10^{-7}$ cases which are qualitatively similar.

To check the effects of Yukawa suppression, we set $y_{\rm env}=1$, which is the maximum value considered in the spherical collapse computation. We then plot $\mathcal{F}$ as a function of dimensionless top-hat radius parameter $y_{\rm h}$ and halo mass, which shows the screening regime and the onset of Yukawa suppression. For large masses and redshifts, screening occurs at larger physical scales while Yukawa suppression occurs at smaller scales. In all cases, the Yukawa suppression is only mildly relevant for $y_{\rm h} \rightarrow y_{\rm env}$ and very large masses. The nPPF best fit value of $p_8$ gives a wrong Yukawa suppression scale, which is likely due to its global fit over all values of $y_{\rm env}$. Similarly, the Erf best fit screening scale, $q_1$, is also underestimated, likely for the same reasons.  

Further, the nPPF shows a good match for the redshift dependence of the screening scale, while the Erf fit does significantly worse. We recall the nPPF uses a theoretically matched power law for this dependence (see \autoref{eq:screeningscale} and \autoref{eq:frpi}), while this dependence is fixed for the Erf case. 

\begin{figure*}
    \centering
    \includegraphics[width=0.95\textwidth]{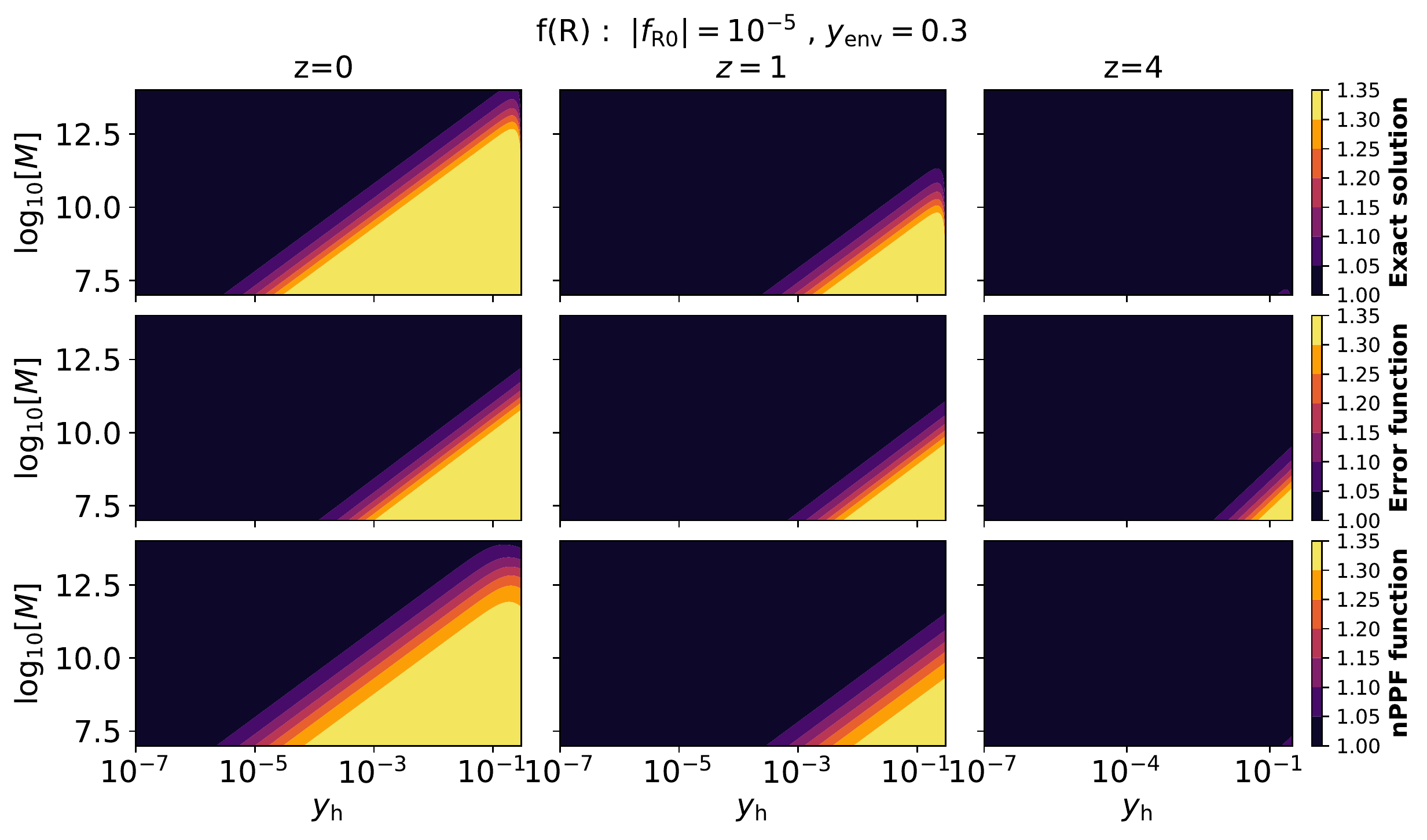}
    \caption{Same as \autoref{fig:f5_Ftest_yenv1} but with $y_{\rm env} = 0.3$.}
    \label{fig:f5_Ftest_yenv03}
\end{figure*}

In \autoref{fig:f5_Ftest_yenv03} we show $1+\mathcal{F}$ for $|f_{\rm R0}| = 10^{-5}$, with $y_{\rm env}=0.3$, again for all cases. We find a good qualitative agreement between the nPPF and exact solutions. On the other hand, the comparisons again show there is an inaccurate redshift dependency in the screening scale of the Erf model, set by $q_1$. This was also seen in the DGP case. Despite this, the flexibility of the model still allows us to produce very accurate results at the power spectrum level (see \autoref{fig:fR_Rtest}) and so we do not feel introducing new freedom is warranted. We leave this issue to be further investigated in future work.

\subsection{A note on notation}\label{app:notation}
We would like to briefly discuss the inconsistency in the notation of previous related publications. The physical top-hat radius is denoted by $R_{\rm TH}$ in \citet{Cataneo:2019fjp, Bose:2020wch}, $r$ in \citet{Carrilho:2021rqo}, $R$ in \citet{Schmidt:2010jr} and $\xi$ in \citet{Lombriser:2013eza}. From the definition of the physical top-hat radius and the conservation of mass $M = 4 \pi \bar{\rho}_\mathrm{m} (\delta+1) R^3_{\rm TH}/3$ the expression for the nonlinear over-density is correctly given in \autoref{eq:nldensity}. Note the corresponding typos in the definition of the nonlinear over-density of \citet{Cataneo:2019fjp, Bose:2020wch, Carrilho:2021rqo} in Equations.~34, B3 and 28 respectively. The connection between the physical top-hat radius and the initial comoving radius $R_{\rm th}$ \footnote{Denoted Rth in the {\tt ReACT} code.} of the over-density is linear $R_{\rm TH}(a_i) = R_i = a_i R_{\rm th}$ initially but then due to the nonlinear evolution of the over-density it becomes $R_{\rm TH}(a) = y a R_{\rm th}$. This nonlinear evolution is encoded in the nonlinear scale factor $ y a$ with $y$ given in \autoref{eq:ydef}. Note that the expression for $\mathcal{F}$ in $f(R)$ gravity in Equation A2 of \citet{Cataneo:2019fjp} and Equation.~C15 of \citet{Bose:2020wch} is taken from \citet{Lombriser:2013eza} and includes $R_{\rm TH}$ which should be replaced by $R_{\rm th} = R_i/a_i$. While  $\mathcal{F}$ in nDGP model from \citet{Schmidt:2010jr} is correctly given in these ReACT papers.

\bsp	
\label{lastpage}
\end{document}